\documentclass[aip,jcp,reprint]{revtex4-2}
\usepackage{amsmath}
\usepackage{bm}
\usepackage{dsfont}
\usepackage[left = 1.35cm, right = 1.35cm, top = 2.2cm, right = 2.2cm]{geometry}

\usepackage{graphicx}

\usepackage{color}

\definecolor{OliveGreen}{rgb}{0.1, 0.4, 0.1}

\usepackage{setspace}
\emergencystretch=\maxdimen
\allowdisplaybreaks

\usepackage[colorlinks=true,citecolor=OliveGreen,linkcolor=OliveGreen,urlcolor=OliveGreen]{hyperref}

\usepackage{mathrsfs}

\begin{document}

\title{Analytical solution for the hydrodynamic resistance of a disk in a compressible fluid layer with odd viscosity on a rigid substrate}%

\author{Abdallah Daddi-Moussa-Ider}%
\email{abdallah.daddi-moussa-ider@open.ac.uk}
\thanks{Corresponding author}
\affiliation{School of Mathematics and Statistics, The Open University, Walton Hall, Milton Keynes MK7 6AA, United Kingdom}

\author{Andrej Vilfan}%
\email{andrej.vilfan@ijs.si}
\affiliation{Jo\v{z}ef Stefan Institute, 1000 Ljubljana, Slovenia}

\author{Yuto Hosaka}%
\email{yuto.hosaka@ds.mpg.de}
\affiliation{Max Planck Institute for Dynamics and Self-Organization (MPI-DS), Am Fa\ss berg 17, 37077 G\"{o}ttingen, Germany}
\date{\today}%

\begin{abstract}
  Chiral active fluids can exhibit odd viscosity, a property that
  breaks the time-reversal and parity symmetries. Here, we examine the
  hydrodynamic flows of a rigid disk moving in a compressible 2D fluid
  layer with odd viscosity, supported by a thin lubrication layer of a
  conventional fluid. Using the 2D Green's function in Fourier space,
  we derive an exact analytical solution for the flow around a disk of
  arbitrary size, as well as its resistance matrix. The resulting
  resistance coefficients break the Onsager reciprocity, but satisfy
  the Onsager-Casimir reciprocity to any order in odd viscosity.
\end{abstract}

\maketitle

\section{Introduction}

Active matter represents a class of systems composed of motile agents that consume energy from their environment, thereby enabling sustained dynamics and activity.\cite{marchetti2013hydrodynamics,bechinger2016active} 
Given its inherently non-equilibrium nature, such systems display complex dynamics, including diverse propulsion mechanisms and collective and motility-induced behaviors across a wide range of scales, from the nanometer to macroscopic scales.\cite{cates2015motility, digregorio2018full, paoluzzi2022motility, sharan2023pair} 
Although both living and synthetic systems of active particles have been extensively investigated in aqueous environments, the suspending fluids themselves can exhibit distinctive active properties by breaking symmetry.\cite{denk2020pattern, tjhung2012spontaneous,shankar2022topological} 
This can result in suspended particles with transport properties that differ from those observed in ordinary fluids.

Among the various types of active fluids, chiral active systems have recently received significant attention.\cite{furthauer2012active, liebchen2022chiral, hosaka2022nonequilibrium, fruchart2023odd, mecke2024emergent, caprini2024self} 
Examples include colloidal suspensions of externally actuated rotating particles~\cite{soni2019odd, lopez2022chirality, zhao2021,mecke2023simultaneous, chen2024self} or media composed of chiral gears.\cite{zhao2022odd} 
These chiral systems can also be realized using biologically-inspired synthetic media composed of spinning constituents.\cite{markovich2021} 
An interesting feature of such fluids is that they are characterized by a non-dissipative transport coefficient called \textit{odd viscosity},\cite{avron1998, banerjee2017} unlike traditional dissipative viscosities. 
These dissipationless and reactive properties of these fluids lead to a rich range of phenomena including surface modes,\cite{souslov2019topological} nonreciprocal hydrodynamic responses, and the notable effects on turbulent cascades.\cite{de2024pattern,chen2024odd}

The important hydrodynamic consequence for transport phenomena in chiral fluids is the existence of a transverse force perpendicular to direction of motion of a particle, namely the lift force.\cite{hosaka2021nonreciprocal, hosaka2021hydrodynamic, lier2023lift, everts2024dissipative, lier2024slip, khain2024trading} 
This is akin to the Magnus force that acts on a spinning object moving through a fluid.\cite{reichhardt2022, lou2022odd}
Comparable transverse or spiraling motions induced by odd viscosity have also been observed for force- and torque-free microswimmers with area-changing properties,\cite{lapa2014} active surface traction,\cite{hosaka2023lorentz} and effective slip velocity.\cite{hosaka2024chirotactic}

Extensive theoretical work has been devoted to developing the hydrodynamics of fluids with odd viscosity at zero Reynolds number where inertial effects are negligible relative to viscous effects. Notably, in two-dimensional (2D) fluids, odd viscosity is compatible with isotropy,\cite{avron1998} and this simplicity has driven a comprehensive study of 2D systems exhibiting odd viscosity. Since fluid chirality introduces contributions from odd viscosity to the classical Stokes equation, it is reasonable to expect odd viscosity to affect the material transport of microparticles. However, in 2D incompressible flows, isotropy-compatible odd viscosity is known to modify only the pressure field in the Stokes equation, leaving the viscous term unchanged.\cite{avron1998} Consequently, the corresponding Green's function remains the same as in classical fluids.\cite{hosaka2021nonreciprocal, khain2022}
Furthermore, it has been shown that as long as the boundary conditions include only the velocity field alone (e.g., no-slip boundary conditions), odd viscosity does not even affect the net force and torque acting on rigid objects of any shape.\cite{ganeshan2017} 

These vanishing effects on the flow field and the resulting force on 2D objects make odd viscosity very elusive, meaning that this transport coefficient cannot be measured experimentally in simple setups.
Several attempts have been made to circumvent this limitation in the incompressible regime. 
For instance, objects with other boundary conditions, such as a deforming bubble with free-surface conditions,\cite{lapa2014, ganeshan2017} translating liquid drops,\cite{hosaka2021hydrodynamic, jia2022incompressible} or a disk with surface slip velocity~\cite{lier2024slip} have been considered and it has been found that the resulting flow fields, as well as the force acting on them are affected by this viscosity.

To overcome the limitations imposed by incompressibility, an alternative to introducing anisotropic odd viscosities~\cite{epstein2020, souslov2020} is to relax the incompressibility condition, allowing mass transport into the surrounding environment, such as a 3D fluid in contact with the 2D layer.
Building on the theory of a compressible thin fluid layer with a soluble surfactant monolayer,\cite{barentin1999} Hosaka \textit{et al.}\ developed a 2D system characterized by odd viscosity and demonstrated that it generates antisymmetric components in the Green’s function, in contrast to the symmetric behavior observed under incompressibility.\cite{hosaka2021nonreciprocal} This model allows even rigid objects to experience lift forces, reflecting the chiral properties of the fluid.

Theoretical models have further explored passive tracer particles in compressible fluids with instantaneous mass relaxation,\cite{hosaka2021nonreciprocal} finite relaxation times,\cite{lier2023lift} or odd visco-elasticity.\cite{duclut2024probe} Studies on active particles, including hydrodynamic force dipoles and linear swimmers, have also been conducted.\cite{hosaka2023pair, hosaka2023hydrodynamics} Nevertheless, despite these extensive theoretical studies, exact analytical solutions of flow problems involving finite-sized particles in fluids of odd viscosity remain very scarce. Apart from problems where odd viscosity does not affect the flow, such as incompressible 2D models with a no-slip boundary \cite{ganeshan2017} or axisymmetric rotation in 3D fluids \cite{hosaka2024chirotactic}, a notable exact solution has been determined for the motion of a sphere in 3D.\cite{everts2024dissipative} Further exact solutions include a flow around a bubble \cite{ganeshan2017} or a liquid inclusion in a 2D odd-viscous fluid.\cite{hosaka2021hydrodynamic}

In this work, we examine the hydrodynamics of a rigid disk moving in a compressible 2D fluid layer, supported by a thin lubrication layer. We solve this problem using the 2D Green’s function, which captures the coupling between the 2D fluid and the underlying 3D fluid. We derive an exact analytical solution for the flow around a disk of arbitrary size, as well as its resistance matrix.

\section{Odd viscosity in 2D fluids}

We begin by reviewing the concept of odd viscosity and its effect on the stress tensor in 2D fluids.\cite{avron1998, epstein2020} 
Let the viscous strain rate tensor be defined as $E_{k\ell} = (\partial_k v_\ell + \partial_\ell v_k)/2$, where $\bm{v}$ represents the 2D~velocity field. The general linear relationship between $E_{k\ell}$ and the fluid stress tensor $\sigma_{ij}$ is expressed as
$\sigma_{ij} =  \eta_{ijk\ell}E_{k\ell}
$ where $\eta_{ijk\ell}$ is the fourth-rank viscosity tensor and invariant under $i\leftrightarrow j$ following from the symmetric stress tensor $\sigma_{ij} = \sigma_{ji}$. In this case, any fourth-rank tensor can be constructed using the Pauli matrices.\cite{avron1998, lapa2014} 
By enforcing rotational invariance around the normal to the 2D fluid layer, the general form of the 2D~viscosity tensor can be expressed as~\cite{hosaka2021nonreciprocal}
\begin{align}
\eta_{i j k\ell} &=
\eta_{\rm D}\delta_{i j}\delta_{k\ell}
+\eta_{\rm S}\left(\delta_{i k} \delta_{j \ell}+\delta_{i \ell} \delta_{j k}-\delta_{i j}\delta_{k\ell}\right)\nonumber\\
&\quad+\tfrac{1}{2} \, \eta_{\rm O}
\left(\epsilon_{i k} \delta_{j \ell}+\epsilon_{j \ell} \delta_{i k}
+\epsilon_{i \ell} \delta_{j k}+\epsilon_{j k} \delta_{i \ell}
\right),
\label{eq:eta}
\end{align}
where $\eta_{\rm D}, \eta_{\rm S}$, and $\eta_{\rm O}$ are 2D dilatational, shear, and odd viscosities, respectively,
$\delta_{ij}$ is the Kronecker delta, and~$\epsilon_{ij}$ is the 2D Levi-Civita tensor with $\epsilon_{xx}=\epsilon_{yy}=0$ and $\epsilon_{xy}=-\epsilon_{yx}=1$.
The viscosity tensor with odd viscosity is antisymmetric under the exchange of indices $ij \leftrightarrow k\ell$, leading to its dissipationless and \textit{reactive} nature.\cite{banerjee2017} In contrast, the components associated with the even viscosities, $\eta_{\rm D}$ and $\eta_{\rm S}$, are symmetric under such an exchange. Unlike the always positive even viscosities, the nondissipative $\eta_{\rm O}$ can be either positive or negative, depending on the chirality of the fluid.

\begin{figure}
    \centering
    \includegraphics[width=0.85\linewidth]{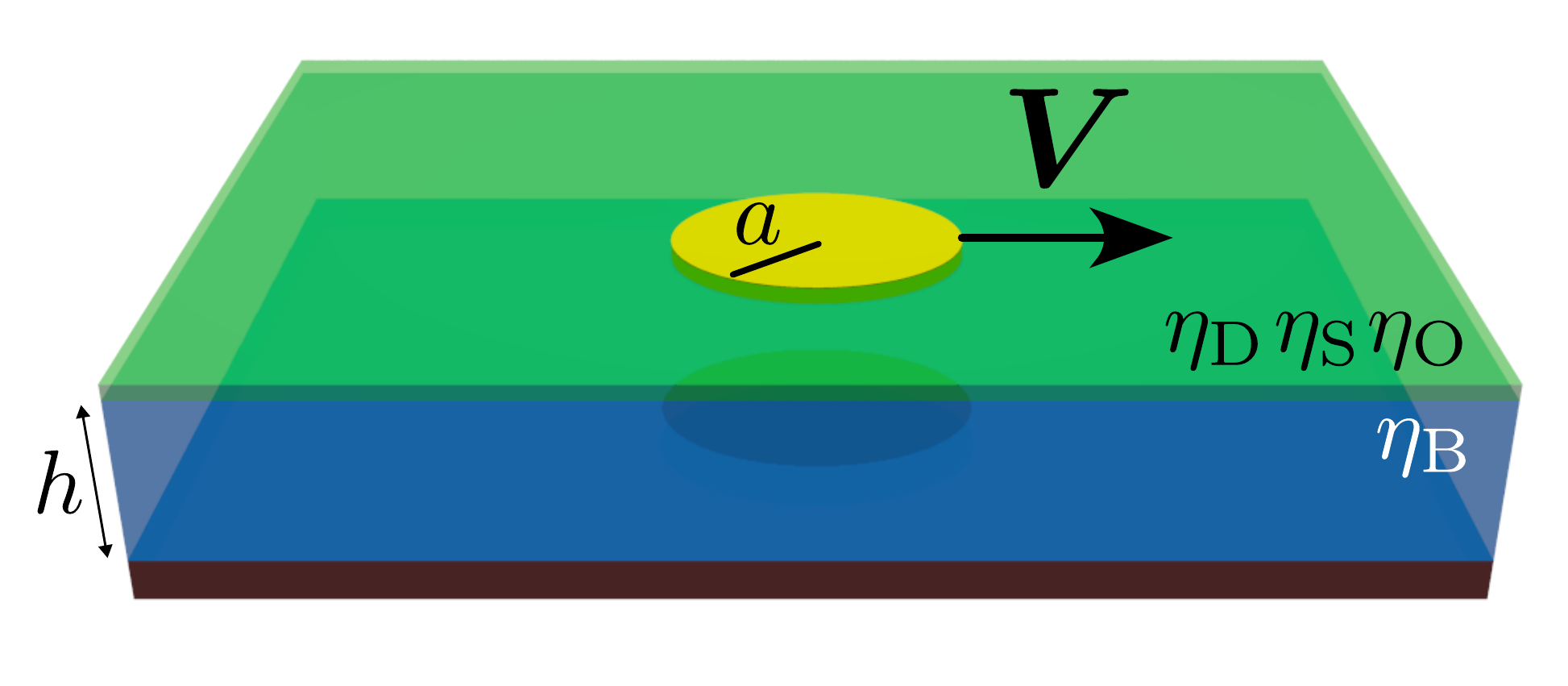} 
    \caption{Schematic illustration of an infinite 2D thin layer of a compressible fluid with 2D dilatational~$\eta_{\rm D}$, shear~$\eta_{\rm S}$, and odd~$\eta_{\rm O}$ viscosities. The fluid layer overlays a 3D bulk fluid medium of a 3D shear viscosity~$\eta_{\rm B}$ and film thickness~$h$, which is supported by a flat and impermeable rigid substrate. The rigid circular disk of radius~$a$ moves laterally within the 2D surface with velocity~$\bm{V}$, subject to the force exerted by the 2D fluid surface on the disk perimeter, as well as the force acting on the disk interior in contact with the underlying 3D fluid. In the sketch, the fluid thickness~$h$ is exaggerated for clarity. In the actual calculations, we assume that~$h$ is much smaller than all lateral dimensions of the system.
    Thus, despite the 3D representation in the sketch, the fluid is effectively treated as a thin layer.\cite{barentin1999, hosaka2023hydrodynamics}
    }
    \label{fig:system}
\end{figure}

Substituting Eq.~\eqref{eq:eta} into the constitutive relation $\sigma_{ij} = \eta_{ijk\ell} E_{k\ell}$, we obtain the stress tensor for a 2D compressible fluid with odd viscosity as
\begin{align}
\sigma_{ij} &=   (\eta_{\rm D}-\eta_{\rm S}) \, \delta_{i j}E_{kk}
+2\eta_{\rm S}E_{ij} 
+\eta_{\rm O}
\left(
\epsilon_{ik}E_{kj}
+
\epsilon_{jk}E_{ki}
\right).
\label{eq:stress}
\end{align}
Due to the presence of the permutation symbol $\boldsymbol{\epsilon}$, it is evident that $\eta_{\rm O}$ induces a transverse or nonreciprocal hydrodynamic response, rotated by $\pi/2$ relative to the standard stress response. In the following, we consider the dynamics of a 2D fluid with the stress tensor given by Eq.~\eqref{eq:stress}, where the fields are described in terms of the 2D position vector $\bm{r} = (x, y)$, ensuring invariance under rotation about the $z$-axis.

\section{Hydrodynamic equations}

We analyze a 2D compressible fluid layer supported from below by an incompressible 3D fluid. The underlying bulk fluid has a 3D shear viscosity $\eta_\mathrm{B}$ and is confined by a flat, rigid wall. The distance between the 2D layer and the wall is fixed at $h$; see Fig.~\ref{fig:system} for an illustration of the system setup. 
The corresponding momentum balance equation for the 2D fluid surface is given by $\boldsymbol{\nabla}_\parallel \cdot \boldsymbol{\sigma} + \bm{f}_{\rm B} + \bm{f} = \mathbf{0}$, where $\boldsymbol{\nabla}_\parallel = (\partial_x, \partial_y)$ is the 2D gradient operator, $\boldsymbol{\sigma}$ is the stress tensor defined in Eq.~(\ref{eq:stress}), $\bm{f}_{\rm B}$ is the force exerted on the 2D fluid layer by the underlying 3D bulk, and $\bm{f}$ represents any other force density acting on the 2D fluid.

It is important to note that the gradient of the 2D pressure does not appear in the momentum balance equation because we assume that the 2D fluid layer quickly equilibrates with the 3D fluid, resulting in a homogeneous 2D pressure in space, as discussed in Ref.~\onlinecite{barentin1999}. This system can be viewed as a monolayer of soluble surfactants, such as a Gibbs monolayer with amphiphiles that can dissolve into the underlying bulk.\cite{barentin1999}

To determine the force from the underlying bulk 3D fluid, we solve the corresponding hydrodynamic equations to obtain $\bm{f}_{\rm B}$.\cite{barentin1999} The bulk beneath the 2D fluid layer is a 3D fluid, governed by the Stokes equation $\eta_{\rm B} \boldsymbol{\nabla}^2 \bm{u} - \boldsymbol{\nabla} p = \mathbf{0}$ and the incompressibility condition $\boldsymbol{\nabla} \cdot \bm{u} = 0$, where $\boldsymbol{\nabla} = (\partial_x, \partial_y, \partial_z)$ is the 3D gradient operator, and $\bm{u}$ and $p$ are the 3D velocity and pressure fields, respectively. 
Sandwiched between the 2D layer at $z = h$ and the impermeable rigid wall at $z = 0$, the 3D flow satisfies the stick boundary conditions: $\bm{u}(\bm{r}, 0) = \bm{0}$ and $\bm{u}(\bm{r}, h) = \bm{v}(\bm{r})$, where $\bm{v}(\bm{r})$ is the velocity of the 2D fluid. 
By applying the lubrication approximation, valid when the thickness $h$ is much smaller than any lateral dimension of the system, the bulk velocity is given by $\bm{u}(\bm{r}, z) = \left( z^2 - zh\right)/\left( 2 \eta_{\rm B} \right) \boldsymbol{\nabla}_\parallel p + (z/h) \, \bm{v}$.\cite{barentin1999} 
Projecting the traction onto the $(x, y)$-plane yields the force exerted by the 3D bulk on the 2D fluid layer, which is given by $\bm{f}_{\rm B} = -(h/2) \boldsymbol{\nabla}_\parallel p - \left( \eta_{\rm B}/h \right) \, \bm{v}$.\cite{barentin1999} 

We have assumed the viscosities of the surface layer remain unchanged as it is compressed. This assumption holds if material transport between the layer and the bulk (e.g., adsorption/desorption process in the case of a Gibbs monolayer) occurs rapidly compared to the motion of the disk, ensuring that the viscosities stay nearly constant.

Through the momentum balance equation, we find the hydrodynamic equation for 2D compressible fluids as follows~\cite{hosaka2021nonreciprocal}
\begin{align}
\boldsymbol{\eta}
\cdot \boldsymbol{\nabla}_\parallel^2 \bm{v}
+ \eta_\mathrm{D}\boldsymbol{\nabla}_\parallel (\boldsymbol{\nabla}_\parallel\cdot\bm{v}) 
-\frac{h}{2} \, \boldsymbol{\nabla}_\parallel p 
-\frac{\eta_\mathrm{B}}{h} \, \bm{v}
+\bm{f} = \bm{0} \, ,
\label{eq:heq}
\end{align}
with 
\begin{equation}
    \boldsymbol{\eta} = \eta_\mathrm{S} \mathds{1}
+ \eta_\mathrm{O}\boldsymbol{\epsilon} \, .
\end{equation}
Additionally, by integrating the divergence of the 3D bulk fluid velocity $\bm{u}(\bm{r}, z)$ over the lubrication layer ($0 \leq z \leq h$) and requiring it to be zero, we obtain the equation
\begin{equation}
    \boldsymbol{\nabla}_\parallel\cdot\bm{v} = \frac{h^2}{6\eta_\mathrm{B}}\boldsymbol{\nabla}_\parallel^2p \, .
    \label{eq:compressibility}
\end{equation}
For a detailed derivation, see Refs.~\onlinecite{barentin1999, elfring2016surface}. 

The hydrodynamic inverse screening lengths, which characterize the distance beyond which the thin fluid layer transports momentum to the underlying bulk fluid, are defined as
\begin{equation}
\kappa = \sqrt{\frac{\eta_\mathrm{B}}{h \eta_\mathrm{S}}} \, , \qquad 
\lambda = \sqrt{\frac{2\eta_\mathrm{B}}{h\bar{\eta}}} \, ,
\label{kappa_lambda}
\end{equation}
with the mean even viscosity $\bar{\eta}=(\eta_\mathrm{S}+\eta_\mathrm{D})/2$.

\section{Green's function}

The linear hydrodynamic response of the 2D fluid layer is characterized by the Green's function \( \boldsymbol{\mathcal{G}}(\bm{r}) \), which relates the applied force to the resulting induced velocity,
\begin{align}
    \bm{v}(\bm{r})\
    =
    \int \boldsymbol{ \mathcal{G} }(\bm{r}-\bm{r}^\prime)\cdot\bm{f}(\bm{r}^\prime) \, \mathrm{d}^2 \bm{r}^\prime \,.
\end{align}
We use a 2D Fourier transform technique to describe the evolution of hydrodynamic fields, a method that has been extensively applied to solving various flow problems in the low Reynolds number regime~\cite{felderhof2006dynamics, daddi2016hydrodynamic, daddi2018brownian, daddi2018hydrodynamic}.
We determine the Green's function from the hydrodynamic equations~\eqref{eq:heq} and~\eqref{eq:compressibility} in Fourier space by representing the fields in terms of the wavevector \( \bm{k} = (k_x, k_y) \).
We denote Fourier-transformed functions with a tilde.
By introducing two orthogonal unit vectors, \( \bm{k}_\parallel = \left(k_x, k_y\right)/k \) and \( \bm{k}_\perp = \left(-k_y, k_x\right)/k \), where \( k = |\bm{k}| \), we can write the Green's function for the velocity field as~\cite{hosaka2021nonreciprocal}
\begin{equation}
\widetilde{\boldsymbol{ \mathcal{G} }} (\bm{k})
= \frac{\eta_\mathrm{S}\left(k^{2}+\kappa^{2}\right) \mathds{P}_\parallel + 2\bar{\eta}\left(k^{2}+\lambda^{2}\right) \mathds{P}_\perp-\eta_{\mathrm{O}} k^{2} \boldsymbol{\epsilon} }{2\eta_\mathrm{S}\bar{\eta}\left(k^{2}+\kappa^{2}\right)\left(k^{2}+\lambda^{2}\right)+\eta_{\mathrm{O}}^{2} k^{4}} , \label{eq:G}
\end{equation}
with $\mathds{P}_\parallel = \bm{k}_\parallel \, \bm{k}_\parallel$ and $\mathds{P}_\perp = \bm{k}_\perp \,  \bm{k}_\perp$.
Here, we have defined the dimensionless parameters 
\begin{equation}
    \mu = \frac{\eta_\mathrm{O}}{\eta_\mathrm{S} } \, , 
    \qquad
    \xi = \frac{\kappa}{\lambda} =\sqrt{\frac{\bar\eta}{2 \eta_\mathrm{S}}}\, ,
\end{equation}
representing the scaled odd viscosity and friction anisotropy parameters, respectively. Because \( \eta_\mathrm{D} \geq 0 \), $\xi$ can only take values \( \xi \geq 1/2 \).

The Green's function for the pressure field follows from Eq.~\eqref{eq:compressibility} as
\begin{equation}
    \widetilde{\bm{\mathcal{P}}}(\bm{k}) = 
    \frac{6i\kappa^2 \eta_\mathrm{S}}{hk}
    \frac{ \eta_\mathrm{O} k^2 \bm{k}_\perp
    -\eta_\mathrm{S} \left( k^2+\kappa^2 \right)
    \bm{k}_\parallel }{ 2\eta_\mathrm{S}\bar{\eta}\left(k^{2}+\kappa^{2}\right)\left(k^{2}+\lambda^{2}\right)+\eta_{\mathrm{O}}^{2} k^{4} } \, .\label{eq:Gp}
\end{equation}

Using polar coordinates, with \( k_x = k \cos\phi \) and \( k_y = k \sin\phi \), and introducing the substitution \( u = k / \kappa \) to represent the scaled wavenumber, the Green's function can be expressed as
\begin{equation}
\hspace{-0.1cm}
    \widetilde{\boldsymbol{ \mathcal{G} }} (u,\phi) =
    \frac{5\left( \alpha u^2+1 \right) \mathds{1}
    -3\left( \beta u^2+1\right)
    \mathds{R}_\phi
    -2\mu u^2 \boldsymbol{\epsilon}}
    { 2\kappa^2 \eta_\mathrm{S} \big( 4 \left( u^2+1 \right) \left( \xi^2 u^2+1\right) + \mu^2 u^4 \big)} ,
\end{equation}
with the double-angle rotation matrix
\begin{equation}
    \mathds{R}_\phi
    = \begin{pmatrix}
        \cos 2\phi & \sin2\phi \\[3pt]
        \sin2\phi & -\cos2\phi
    \end{pmatrix} ,
\end{equation}
where we have defined the dimensionless numbers 
\begin{subequations}
    \begin{align}
    \alpha &= \frac{1}{5} \left( 4\xi^2+1 \right)= \frac {\eta_\mathrm{D}+2 \eta_\mathrm{S}}{5 \eta_\mathrm{S}} \, , \\
    \beta &= \frac{1}{3} \left( 4\xi^2-1 \right) = \frac {\eta_\mathrm{D}}{3 \eta_\mathrm{S}}
     \, ,
\end{align}
\end{subequations}
such that $\alpha \ge 2/5$ and $\beta \ge 0$.

Likewise the Green's function for the pressure can be written as
\begin{equation}
    \widetilde{\bm{\mathcal{P}}}(u,\phi) = 
    \frac{6i}{\kappa h u}
    \frac{ \mu u^2\, \bm{k}_\perp -(u^2+1) \, \bm{k}_\parallel }{4 \left( u^2+1 \right) \left( \xi^2 u^2+1\right) + \mu^2 u^4} \, ,
    \label{eq:pressure-fourier}
\end{equation}
where, $\bm{k}_\parallel = \left( \cos\phi, \sin\phi \right)$ and~$\bm{k}_\perp = \left( -\sin\phi, \cos\phi\right)$, expressed in terms of the polar angle~$\phi$.

\section{Velocity field of a disk in motion}

We consider the steady translational motion of a circular disk moving laterally in a 2D fluid layer, particularly characterized by an odd viscosity; see Fig.~\ref{fig:system}. The disk has a circular shape with radius \( a \) and moves with a translational velocity \( \bm{V} \).
We denote \( r \) as the radial distance and \( \theta \) as the azimuthal angle in polar coordinates.
We impose a no-slip boundary condition at the disk surface. 
Without loss of generality, we assume the motion occurs along the positive \( x \)-direction, so that \( \bm{V} = V \hat{\bm{e}}_x \). In the following sections, we derive the resulting velocity and pressure fields around the disk by analytically solving the 2D odd Stokes equations \eqref{eq:heq} and~\eqref{eq:compressibility}.

To determine the hydrodynamic fields, we seek the unknown force density that satisfies the velocity boundary conditions.\cite{Daddi-Moussa-Ider_2024_JPCM, daddi2024rotational, daddi2024hydrodynamic_JFM} 
As an Ansatz, we express the force density as a linear superposition of a constant force applied across the entire surface of the disk and an angular-dependent contribution at the disk's perimeter.
Specifically,
\begin{equation}
    \bm{f}({r}, \theta) = 
    \frac{\eta_\mathrm{S} V}{a^2}
    \left( 
    \bm{f}_\Theta ({r}, \theta) 
    + \bm{f}_\delta ({r}, \theta)
    \right) \, , \label{eq:force_real}
\end{equation}
where
\begin{subequations}
    \begin{align}
    \bm{f}_\Theta ({r}, \theta) &= 2\bm{B} \, \Theta(a-{r}) \, , \\[3pt]
    \bm{f}_\delta ({r}, \theta) &= 
    \big( \bm{A}-\bm{B} - \mathds{R}_\theta \cdot \bm{C} \big) \, a \, \delta({r}-a) \, , 
\end{align}
\end{subequations}
wherein $\bm{A}$, $\bm{B}$ and~$\bm{C}$ are unknown 2D~vector coefficients chosen dimensionless to be determined from the boundary conditions.
We have selected this form to achieve a simplified expression in Fourier space, facilitating subsequent derivation.
The angular dependence follows that of the Green's function, containing only  the zeroth and second Fourier mode.
The 2D Fourier transform of Eq.~\eqref{eq:force_real} is obtained as~\cite{baddour2011two}
\begin{equation}
    \widetilde{ \bm{f} }(k,\phi) = 
    2\pi \eta_\mathrm{S} V
    \big(
     \bm{A}  J_0(ka) 
    + \left( \bm{B} + \mathds{R}_\phi \cdot \bm{C} \right) J_2(ka) \big),
    \label{eq:tractionink}
\end{equation} 
with $J_n$ denoting the $n$th-order Bessel function of the first kind.\cite{abramowitz2000handbook}
Denoting the vector \( \bm{\mathcal{X}} = \left( \bm{A}, \bm{B}, \bm{C} \right)^\top \), where $\top$ denotes the transpose operation, as the column vector of six unknown coefficients in the force density, the velocity field is then obtained as
\begin{equation}
    \bm{v}({r}, \theta) = V \big( \bm{\mathcal{V}}_1 ({r}) + \mathds{R}_\theta \cdot \bm{\mathcal{V}}_2 ({r}) \big) \cdot \bm{\mathcal{X}} \, , 
    \label{eq:velocity-general-form}
\end{equation}
where $\bm{\mathcal{V}}_1$ and~$\bm{\mathcal{V}}_2$ are $2\times 6$ matrices which are functions of the radial coordinates~${r}$ given by
\begin{subequations} \label{eq:V1V2}
    \begin{align}
    \bm{\mathcal{V}}_1 &= 
    \left( \bm{G}_{00}^\top~|~\bm{G}_{10}^\top~|~\bm{H}_{10} \right) \, , \\[3pt]
    \bm{\mathcal{V}}_2 &= 
    \left( \bm{H}_{01}~|~\bm{H}_{11}~|~\bm{G}_{11} \right) \, ,
\end{align}
\end{subequations}
with
\begin{subequations}
    \begin{align}
    \bm{G}_{mn} &= \frac{5}{2} \left( \Phi_{mn}^0+\alpha \Phi_{mn}^1 \right) \mathds{1}
    + \mu \, \Phi_{mn}^1 \boldsymbol{\epsilon} \, , \\
    \bm{H}_{mn} &= \frac{3}{2} \left( \Phi_{mn}^0 + \beta \Phi_{mn}^1 \right) \mathds{1} \, .
\end{align}
\end{subequations}

The series functions \( \Phi_{mn}^q \), with \( m, n, q \in \{0, 1\} \), are defined in terms of the convergent improper integrals
\begin{equation}
    \Phi_{mn}^q  = \int_0^\infty
    \frac{u^{2q+1} J_{2m}(bu) J_{2n}({\rho} u) \, \mathrm{d}u}{4 \left( u^2+1 \right) \left( \xi^2 u^2+1\right) + \mu^2 u^4} \, , \label{eq:phi_mn_q}
\end{equation}
with \( b = \kappa a \) and \( \rho = \kappa r \), representing the scaled disk radius and polar distance, respectively.

\subsection{Evaluation of $\Phi_{mn}^q$}

To evaluate this integral analytically, we carry out the integration in complex plane using the residue theorem. We define the function of the complex variable
\begin{equation}
    f(z) = 
    \begin{cases}
        \cfrac{z^{2q+1} H_{2m}^{(1)}(bz) J_{2n}({\rho} z)}{4 \left( z^2+1 \right) \left( \xi^2 z^2+1\right) + \mu^2 z^4} & \text{if } {\rho} \le b \\[3pt]
        \cfrac{z^{2q+1} J_{2m}(bz) H_{2n}^{(1)}({\rho} z)}{4 \left( z^2+1 \right) \left( \xi^2 z^2+1\right) + \mu^2 z^4}  & \text{if } {\rho} \ge b
    \end{cases}
    \label{eq:f_of_z}
\end{equation}
with $H_n^{(1)}$ denoting the $n$th-order Hankel function of the first kind.\cite{abramowitz2000handbook}
It is evident that $f(z)$ is a meromorphic function, as it is analytic everywhere except at a finite number of poles.

We integrate the function \( f(z) \) along a contour consisting of the following segments: the upper side of the branch cut from \( -R \) to \( -\epsilon \), a small clockwise-oriented semicircle of radius \( \epsilon \) centered at the origin, the positive real axis from \( \epsilon \) to \( R \), and the counterclockwise path along the upper half of the circle \( |z| = R \).
We have chosen \( f(z) \) based on whether \( \rho \leq b \) or \( \rho \geq b \) to ensure that the integral over the large semi-circle of radius \( R \) vanishes as \( R \to \infty \).

The poles of the function $f(z)$ defined by Eq.~\eqref{eq:f_of_z} are located at
\( z = iA_\pm \), where 
\begin{equation}
    A_\pm = \left( \frac{2 \left( 1+\xi^2 \pm \delta^{\frac{1}{2}} \right)}{\mu^2 + 4 \xi^2} \right)^\frac{1}{2} 
\end{equation}
and
\begin{equation}
    \delta = \left( 1-\xi^2\right)^2-\mu^2 \, .
\end{equation}
The imaginary part of \( z = iA_\pm \) is always positive, placing the poles in the upper half of the complex plane. For \( \delta \geq 0 \), \( z = iA_\pm \) are purely imaginary, while for \( \delta \leq 0 \), \( A_+ \) and \( A_- \) are complex conjugates. The integral is then determined by the sum of the residues
\begin{equation}
    \Phi_{mn}^q =  \Gamma_{mn}^q(A_-) - \Gamma_{mn}^q(A_+) + K_{mn}^q  \, , \label{eq:phi_mn_q_res}
\end{equation}
where
\begin{equation}
    \Gamma_{mn}^q(z) = \frac{i\pi z^{2q}}{8\delta^\frac{1}{2}}
    \begin{cases}
        (-1)^{n+q} \, H_{2m}^{(1)}(ib z) I_{2n}({\rho} z) & \text{if } \rho \le b \\[3pt]
        (-1)^{m+q} \, I_{2m}(bz)H_{2n}^{(1)}(i{\rho} z) & \text{if } \rho \ge b
    \end{cases}
\end{equation}
with $I_n$ denoting the $n$th-order modified Bessel function of the first kind.\cite{abramowitz2000handbook}
In addition, 
\begin{equation}
    K_{mn}^q = 
    \begin{cases}
        \cfrac{1}{2b^2} \,\, \delta_{m,1} \delta_{n,0}  \delta_{q,0} & \text{if } {\rho} \le b \\[3pt]
        \cfrac{1}{2{\rho}^2} \,\, \delta_{m,0} \delta_{n,1} \delta_{q,0} & \text{if } {\rho} \ge b
    \end{cases}
    \label{eq:K_mn_q_res}
\end{equation}
which stems from the integration over half a circle around the origin and vanishes except for 
$(m,n,q)=(1,0,0)$ when ${\rho} \le b$ and
$(m,n,q)=(0,1,0)$ when ${\rho} \ge b$.
It is worthwhile to note the symmetry relation $\Phi_{mn}^q(b, {\rho}) = \Phi_{nm}^q({\rho},b)$.

\subsection{Velocity within the disk}

To impose the no-slip boundary conditions at the surface of the disk, we need the expression for the velocity field inside the disk for \( \rho \le b \).
By making use of Eqs.~\eqref{eq:phi_mn_q_res}–\eqref{eq:K_mn_q_res}, the expressions for \( \bm{\mathcal{V}}_1 \) and \( \bm{\mathcal{V}}_2 \) defining the velocity field in Eqs.~\eqref{eq:V1V2} are
\begin{subequations}
    \begin{align}
    \bm{\mathcal{V}}_1 &= 
    \bm{\mathcal{W}}_1^+ I_0({\rho} A_+)
    +\bm{\mathcal{W}}_1^- I_0({\rho} A_-)
    + \bm{\mathcal{C}}_1 \, , \\[3pt]
    \bm{\mathcal{V}}_2 &= 
    \bm{\mathcal{W}}_2^+ I_2({\rho} A_+)
    +\bm{\mathcal{W}}_2^- I_2({\rho} A_-) \, ,
\end{align}\label{eq:V1V2-inside}%
\end{subequations}%
where 
\begin{equation}
     \bm{\mathcal{C}}_1 = 
    \frac{1}{4b^2} \,
    \big( ~\bm{0}~\big\rvert~5 \mathds{1}~\big\rvert~3 \mathds{1}~ \big) \, .
\end{equation}
In addition,
\begin{subequations}
 \begin{align}
    \bm{\mathcal{W}}_1^\pm &= 
    \pm \frac{i\pi}{8\delta^\frac{1}{2}}  \left( \left(\boldsymbol{\Pi}_0^\pm\right)^\top~\Big\rvert~\left(\boldsymbol{\Pi}_2^\pm\right)^\top~\Big\rvert~\boldsymbol{{\Sigma}}_2^\pm \right) , \\[3pt]
    \bm{\mathcal{W}}_2^\pm &= 
    \mp \frac{i\pi}{8\delta^\frac{1}{2}}  \big( ~\boldsymbol{{\Sigma}}_0^\pm~\big\rvert~\boldsymbol{{\Sigma}}_2^\pm~\big\rvert~\boldsymbol{\Pi}_2^\pm~ \big) \, .
\end{align}
\end{subequations}
Here, we have defined the abbreviations
\begin{subequations}
    \begin{align}
    \boldsymbol{\Pi}_n^\pm &= 
    \left( 
    ~\frac{5}{2} \, \left( \alpha A_\pm^2-1 \right)  \mathds{1}
    + \mu A_\pm^2 \, \boldsymbol{\epsilon}~
    \right) H_n^{(1)}(ibA_\pm) \, , \\[3pt]
    \boldsymbol{{\Sigma}}_n^\pm &= 
    \frac{3}{2} \, \left( \beta A_\pm^2-1 \right)
    H_n^{(1)}(ibA_\pm) \, \mathds{1} .
\end{align}
\end{subequations}

\subsection{Velocity outside the disk}

The key result of this paper is the determination of the velocity field outside the disk.
For ${\rho} \ge b$, the corresponding expressions of $\bm{\mathcal{V}}_1$ and~$\bm{\mathcal{V}}_2$ as defined by Eqs.~\eqref{eq:V1V2} can likewise be obtained.
Using Eqs.~\eqref{eq:phi_mn_q_res}–\eqref{eq:K_mn_q_res} providing explicit expressions of the series integrals $\Phi_{mn}^q$, we obtain
\begin{subequations}
    \begin{align}
    \bm{\mathcal{V}}_1 &= 
    \bm{\mathcal{V}}_1^+ H_0^{(1)}(i{\rho} A_+)
    +\bm{\mathcal{V}}_1^- H_0^{(1)}(i{\rho} A_-) \, , \\[3pt]
    \bm{\mathcal{V}}_2 &= 
    \bm{\mathcal{V}}_2^+ H_2^{(1)}(i{\rho} A_+)
    +\bm{\mathcal{V}}_2^- H_2^{(1)}(i{\rho} A_-) + \bm{\mathcal{C}}_2\, ,
\end{align}\label{eq:V1V2-outside}%
\end{subequations}%
where 
\begin{equation}
     \bm{\mathcal{C}}_2 = 
    \frac{3}{4{\rho}^2} \,
    \big( \mathds{1}~\big\rvert~\bm{0}~\big\rvert~\bm{0}~\big) \, .
\end{equation}
In addition,
\begin{subequations}
 \begin{align}
    \bm{\mathcal{V}}_1^\pm &= 
    \frac{i\pi}{8\delta^\frac{1}{2}} 
    \left( \pm \left(\boldsymbol{\Lambda}_0^\pm\right)^\top~\Big\rvert~\mp\left(\boldsymbol{\Lambda}_2^\pm\right)^\top~\Big\rvert~\mp\boldsymbol{\Psi}_2^\pm \right) , \\[3pt]
    \bm{\mathcal{V}}_2^\pm &= 
    \frac{i\pi}{8\delta^\frac{1}{2}} 
    \big( ~\pm\boldsymbol{\Psi}_0^\pm~\big\rvert~\mp\boldsymbol{\Psi}_2^\pm~\big\rvert~\mp\boldsymbol{\Lambda}_2^\pm~ \big) \, .
\end{align}
\end{subequations}
Here, we have defined
\begin{subequations}
    \begin{align}
    \boldsymbol{\Lambda}_n^\pm &= 
    \left( 
    ~\frac{5}{2} \, \left( \alpha A_\pm^2-1 \right)  \mathds{1}
    + \mu A_\pm^2 \, \boldsymbol{\epsilon}~
    \right) I_n(bA_\pm) \, , \\[3pt]
    \boldsymbol{\Psi}_n^\pm &= 
    \frac{3}{2} \, \left( \beta A_\pm^2-1 \right)
    I_n(bA_\pm) \, \mathds{1} \, .
\end{align}
\end{subequations}

\begin{figure}
    \centering
    \includegraphics[width=0.85\linewidth]{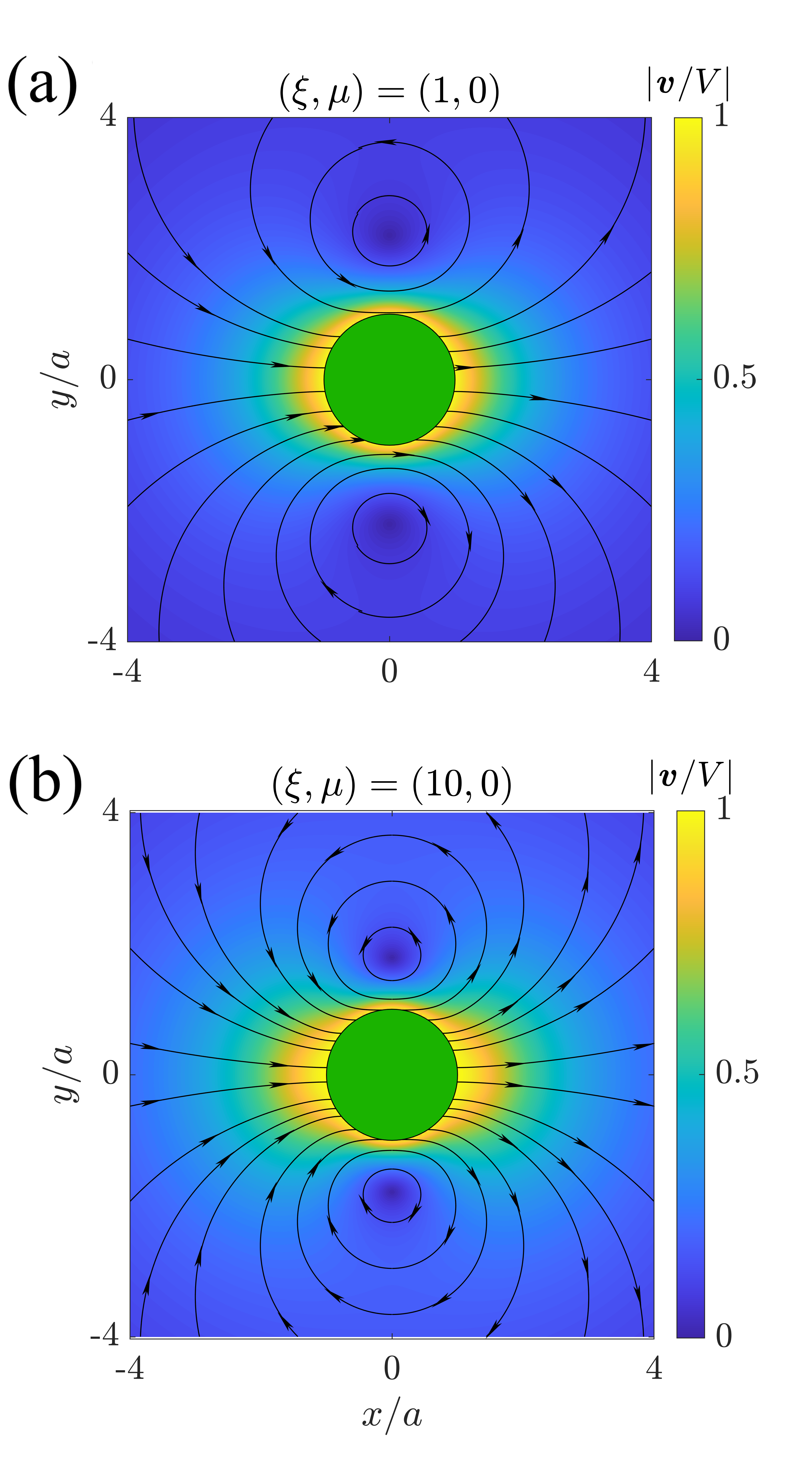} 
    \caption{
    Streamlines of the velocity field \( \bm{v} \) induced by the motion of a disk with velocity \( \bm{V} = (V, 0) \) and radius \( \kappa a = 1 \), for (a) the dilatational-to-shear viscosity ratio \( \xi = \kappa / \lambda = 1 \) and (b) \( \xi = 10 \), while keeping the odd viscosity at zero (\( \mu = 0 \)). The colorbar represents the magnitude of the induced velocity, scaled by the translation velocity \( V \).
    }
    \label{fig:varying_xi}
\end{figure}

\subsection{Solution for the force density}

The coefficients of the force density, represented by the six-dimensional vector \( \bm{\mathcal{X}} \), can now be determined by applying the velocity boundary condition at the inner surface of the disk. The no-slip boundary conditions imply that
\begin{equation}
    \bm{\mathcal{C}}_1 \cdot \bm{\mathcal{X}} = \hat{\bm{e}}_x \,, \qquad
    \bm{\mathcal{W}}_n^\pm \cdot \bm{\mathcal{X}} = \bm{0} \, , \quad n \in \{1,2\} \, .
    \label{eq:System10eqns}
\end{equation}
This yields a set of 10 equations, of which 6 are linearly independent and define the unknown force density coefficients \( \bm{\mathcal{X}} \). Thus, it suffices to consider the system of 6 linearly independent equations, e.g.\ for $n=2$. 
The final results can be expressed as
\begin{subequations}
    \begin{align}
        \bm{A} &= \frac{8b^2}{D} \, \delta^\frac{1}{2} B_2 
        \begin{pmatrix}
            c_- G_- - c_+ G_+ \\[3pt]
            4\mu \left( G_+ - G_- \right)
        \end{pmatrix} , \label{eq:A} \\[7pt]
        \bm{B} &= \frac{b^2}{D}
        \begin{pmatrix}
            4\left( g_+ G_+^2 + g_- G_-^2 + v B_0 B_2 \right) \\[3pt]
            3\mu \left( G_+-G_- \right) \left( d_+ G_+ - d_- G_- \right)
        \end{pmatrix} , \label{eq:B} \\[7pt]
        \bm{C} &= \frac{b^2}{D}
        \left( G_+-G_- \right)
        \begin{pmatrix}
            4\delta^\frac{1}{2}
            \left( d_+G_++d_-G_- \right) \\[3pt]
            5\mu \left( d_-G_--d_+G_+ \right)
        \end{pmatrix} , \label{eq:C}
    \end{align}\label{eq:ABC}
\end{subequations}
with the denominator 
\begin{equation}
    D = h_+ G_+^2 + h_- G_-^2 + w B_0 B_2 \, .
\end{equation}
Here, we have defined the abbreviations 
\begin{subequations}
    \begin{align}
    B_n &= H_n^{(1)}(ibA_+) H_n^{(1)}(ibA_-) \, , \\
    G_\pm &= H_0^{(1)}(ibA_\pm) H_2^{(1)}(ibA_\mp) \, .
\end{align}
\end{subequations}
In addition, $c_\pm = 3\left( \xi^2-1\right) \pm 5\delta^\frac{1}{2}$, $d_\pm = 5\left( \xi^2-1\right) \pm 3 \delta^\frac{1}{2}$, $h_\pm =  c_\pm^2 + 16\mu^2$, $g_\pm = d_\pm \left( \xi^2-1\right)$, $w = 2\left( 16\delta-25\mu^2\right)$, and $v = 10 \left( \delta-\mu^2 \right)$.


In Fig.~\ref{fig:varying_xi}, we present the velocity field induced by a disk translating along the positive \( x \)-direction for dilatational-to-shear viscosity ratios \( \xi = 1 \) and \( 10 \), with odd viscosity set to zero. The flow field exhibits symmetry with respect to the \( x \)-axis in the absence of odd viscosity. Circulating flows, indicative of the momentum leakage to the bulk fluid, mark the transition between purely 2D hydrodynamics and 3D effects. As the parameter \( \xi \) increases, corresponding to a higher dilatational viscosity or lower shear viscosity, the velocity magnitude at the front and rear of the disk also increases.

\begin{figure}
    \centering
    \includegraphics[width=0.85\linewidth]{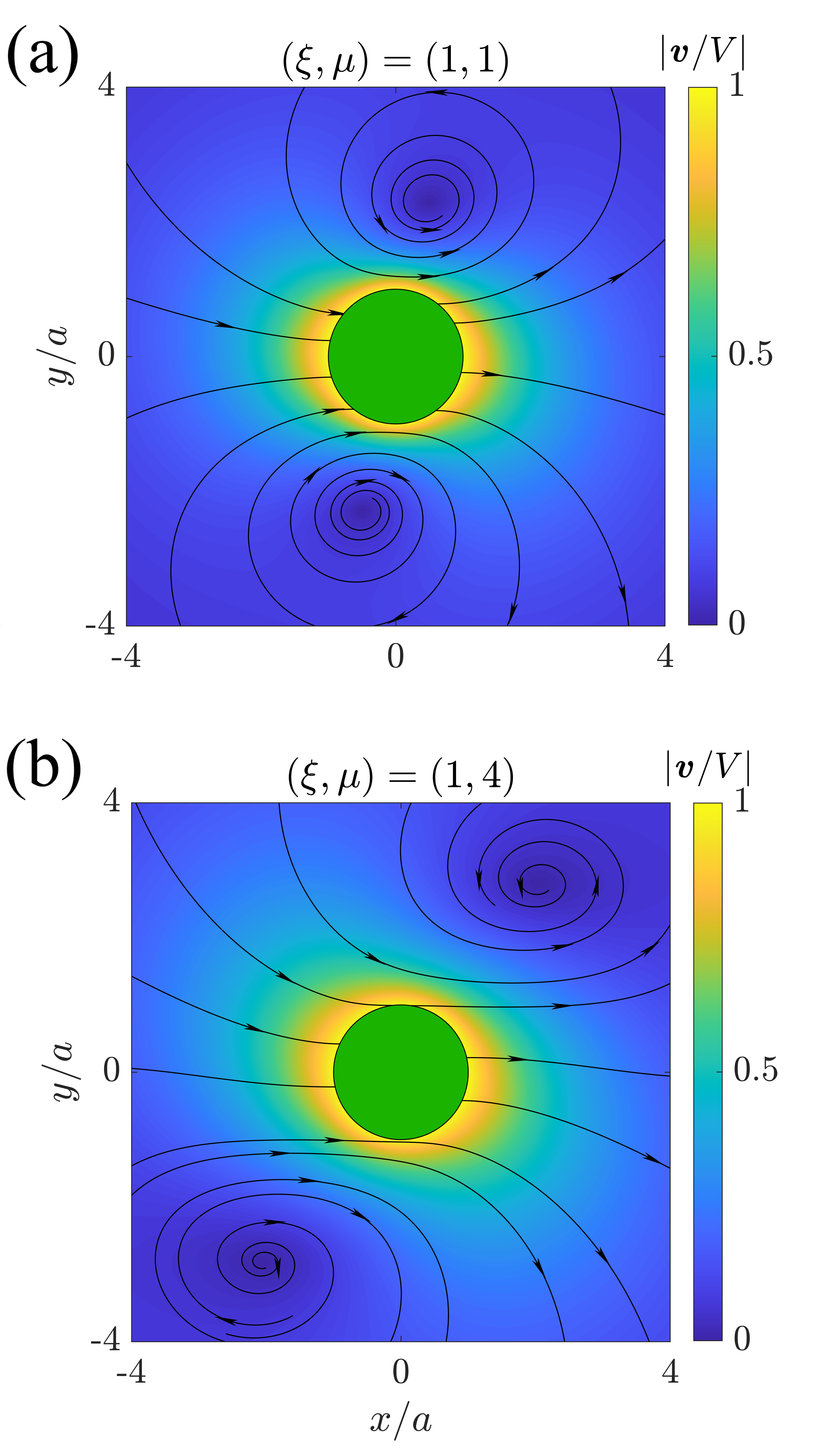} 
    \caption{
    Streamlines and magnitude of the scaled velocity field \( \bm{v} \) induced by the motion of a disk with radius \( \kappa a = 1 \) moving with velocity \( \bm{V} = (V, 0) \) are shown for (a) the odd-to-shear viscosity ratio \( \mu = \eta_{\rm O} / \eta_{\rm S} = 1 \) and (b) \( \mu = 4 \), while keeping \( \xi = 1 \).
   }
    \label{fig:varying_mu}
\end{figure}

When the odd viscosity is nonzero \( (\mu \neq 0) \), a nonreciprocal hydrodynamic response develops transverse to the \( y \)-axis, breaking the mirror symmetry across the $(x,z)$-plane, as shown in Fig.~\ref{fig:varying_mu}. As \( \mu \) increases, the centers of the circulating flows shift toward larger \( x \)-values, indicating an effective increase in the hydrodynamic screening length determined by \( \xi = \kappa / \lambda \). For negative values of the odd viscosity (\( \mu < 0 \)), the corresponding streamlines can be obtained by applying the mirror reflection across the $(y,z)$-plane, which maps $(x, y) \to (-x, y)$ onto the velocity field with a positive odd viscosity of equal magnitude.

\section{Pressure field}

The pressure field can be determined by applying the inverse Fourier transform to Eq.~\eqref{eq:pressure-fourier}. This requires evaluating another set of improper integrals, which can also be solved using the method of residues. However, a more efficient approach is to seek the solution directly in real space, utilizing the known solution for the velocity field. We anticipate that the pressure field will take the following form
\begin{equation}
    p = P \, \big( \bm{\mathcal{P}}_1(\rho) \cos\theta + \bm{\mathcal{P}}_2(\rho) \sin\theta \big) \cdot \bm{\mathcal{X}} \, , 
    \label{eq:pressure-general-form}
\end{equation}
Here, 
\begin{equation}
    P = \frac{\kappa \eta_\mathrm{S} V}{h} 
\end{equation}
denotes the dimension of the pressure field. Additionally, \( \bm{\mathcal{P}}_1 \) and \( \bm{\mathcal{P}}_2 \) are two unknown \( 1 \times 6 \) matrices, which can be determined by solving the corresponding ordinary differential equations in real space.

By substituting Eqs.~\eqref{eq:velocity-general-form} and~\eqref{eq:pressure-general-form} into the compressibility equation~\eqref{eq:compressibility} in polar coordinates, we derive a differential equation in the radial coordinate
\begin{equation}
    \frac{1}{6} \,\,  \mathscr{D} \, \bm{\mathcal{P}} 
       = 
    \frac{\mathrm{d}}{\mathrm{d} \rho}
    \left( \bm{\mathcal{V}}_1+\bm{\mathcal{V}}_2 \right)
    +\frac{2}{\rho} \, \bm{\mathcal{V}}_2  \, ,
\end{equation}
with the differential operator
\begin{equation}
    \mathscr{D} = \frac{\mathrm{d}^2}{\mathrm{d} \rho^2} + \frac{1}{\rho}\frac{\mathrm{d}}{\mathrm{d} \rho} - \frac{1}{\rho^2} \, ,
\end{equation}
and the unknown vector
\begin{equation}
    \bm{\mathcal{P}} = \begin{pmatrix}
        \bm{\mathcal{P}}_1 \\
        \bm{\mathcal{P}}_2
    \end{pmatrix} .
\end{equation}

We have demonstrated that the expressions for $\bm{\mathcal{V}}_1$ and $\bm{\mathcal{V}}_2$ take different forms depending on whether the evaluation point lies inside or outside the disk. For details, see Eqs.~\eqref{eq:V1V2-inside} and \eqref{eq:V1V2-outside}, which provide the corresponding expressions of these matrices in the interior and exterior of the disk, respectively.

For the fluid domain outside the disk, where ${\rho} \ge b$, the solution, which must remain regular at infinity, is given by
\begin{equation}
    \bm{\mathcal{P}} = 
    \bm{\mathcal{C}}_3 
    + \bm{\mathcal{H}}^+
    + \bm{\mathcal{H}}^-
\end{equation}
where
\begin{equation}
     \bm{\mathcal{C}}_3 = 
    \frac{3}{2{\rho}} \,
    \big( \mathds{1}~\big\rvert~\bm{0}~\big\rvert~\bm{0}~\big) \, ,
\end{equation}
and
\begin{equation}
    \bm{\mathcal{H}}^\pm = 
    6i \, \left( \bm{\mathcal{V}}_2^\pm - \bm{\mathcal{V}}_1^\pm \right) \frac{ H_1^{(1)}(i{\rho} A_\pm) }{A_\pm} \, .
\end{equation}
Note that \(\bm{\mathcal{C}}_3\) is obtained by solving Eq.~\eqref{eq:heq}, as the source term, which decays as \(1/\rho\), naturally satisfies the incompressibility condition in Eq.~\eqref{eq:compressibility}.

The Laplacian of the pressure inside the disk vanishes for ${\rho} \le b$, resulting in a solution that is regular at the origin, given by
\begin{equation}
    \bm{\mathcal{P}} = \frac{3{\rho} }{2b^2} \,
    \big( ~\bm{0}~\big\rvert~\mathds{1}~\big\rvert~-\mathds{1}~ \big) \, .
\end{equation}
It can be easily verified that both the pressure field and its first derivative are continuous at ${\rho} = b$.

\begin{figure}
    \centering
    \includegraphics[width=0.85\linewidth]{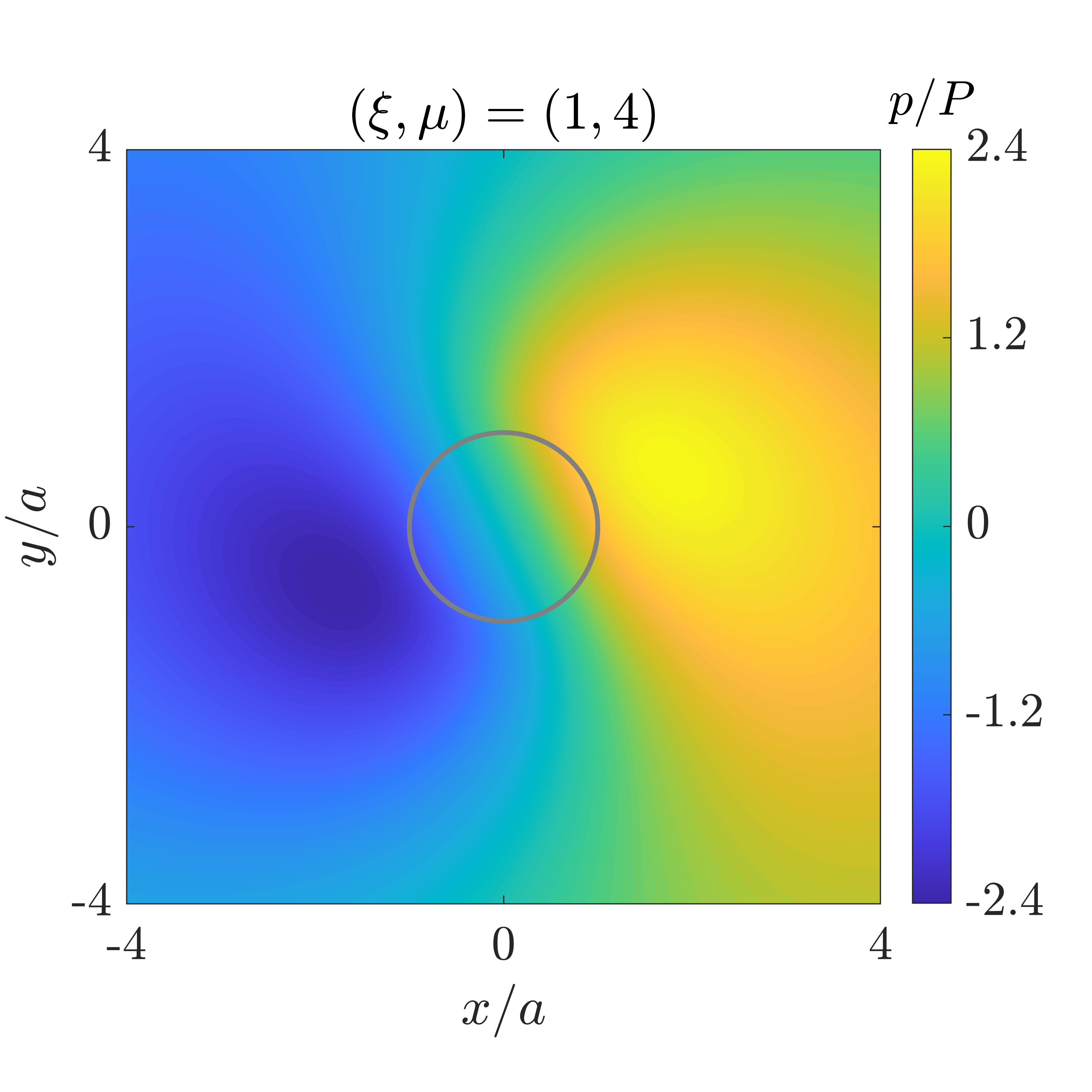}
    \caption{
      Pressure field in the thin bulk fluid induced by the translational motion of a disk moving with velocity \( \bm{V} = (V, 0) \), for the odd-to-shear viscosity ratio \( \mu = \eta_{\rm O}/\eta_{\rm S} = 4 \), \( \xi = 1 \), and \( b=\kappa a = 1 \). The pressure is scaled by \( P = \kappa \eta_{\rm S} V / h \).
    The gray circle represents the perimeter of the disk.}
    \label{figpress:varying_mu}
\end{figure}

In Fig.~\ref{figpress:varying_mu}, we present an exemplary contour plot of the scaled pressure field for \( (\xi, \mu) = (1, 4) \) and \( \kappa a = 1 \). The fluid near the front of the disk exhibits positive pressure, while the pressure in the rear is negative. The effect of odd viscosity is evident in the shifting of the high- and low-pressure regions away from the axis of motion. The pressure field clearly shows a smooth transition between the inner and outer disk, demonstrating the robustness of our analytical approach.

\section{Resistance coefficients}

We now examine the effect of a substrate-supported odd fluid layer on the translational dynamics of a circular disk by calculating the resistance coefficient. Using our framework, the resistance force exerted on the disk by the surrounding fluid is obtained by averaging the force density over the surface of the disk.
Specifically we find from Eq.~\eqref{eq:tractionink}
\begin{equation}
    \bm{F}_\mathrm{R} = -\widetilde{\bm{f}}(k=0) = -2\pi \eta_\mathrm{S} V \bm{A} \, ,
\end{equation}
where the expression for $\bm{A}$ is given in Eq.~\eqref{eq:A}.

We define the resistance $\Gamma_\parallel$ and the lift coefficient $\Gamma_\perp$, which link the velocity of the disk to the hydrodynamic force exerted on it, as
\begin{equation}
    \begin{pmatrix}
        \Gamma_\parallel \\
        \Gamma_\perp
    \end{pmatrix}
    = 
    - \bm{F}_\mathrm{R} \big/ V  = 
    2\pi \eta_\mathrm{S} \bm{A} \, .
\end{equation}
In the limit as $\mu \to 0$, the lift coefficient vanishes, leaving only a resistance force that opposes the particle's motion.
We confirm that a translating disk does not experience any torque, as required by the symmetry of chiral fluids.\cite{khain2024trading}

\begin{figure}
    \centering
    \includegraphics[width=0.85\linewidth]{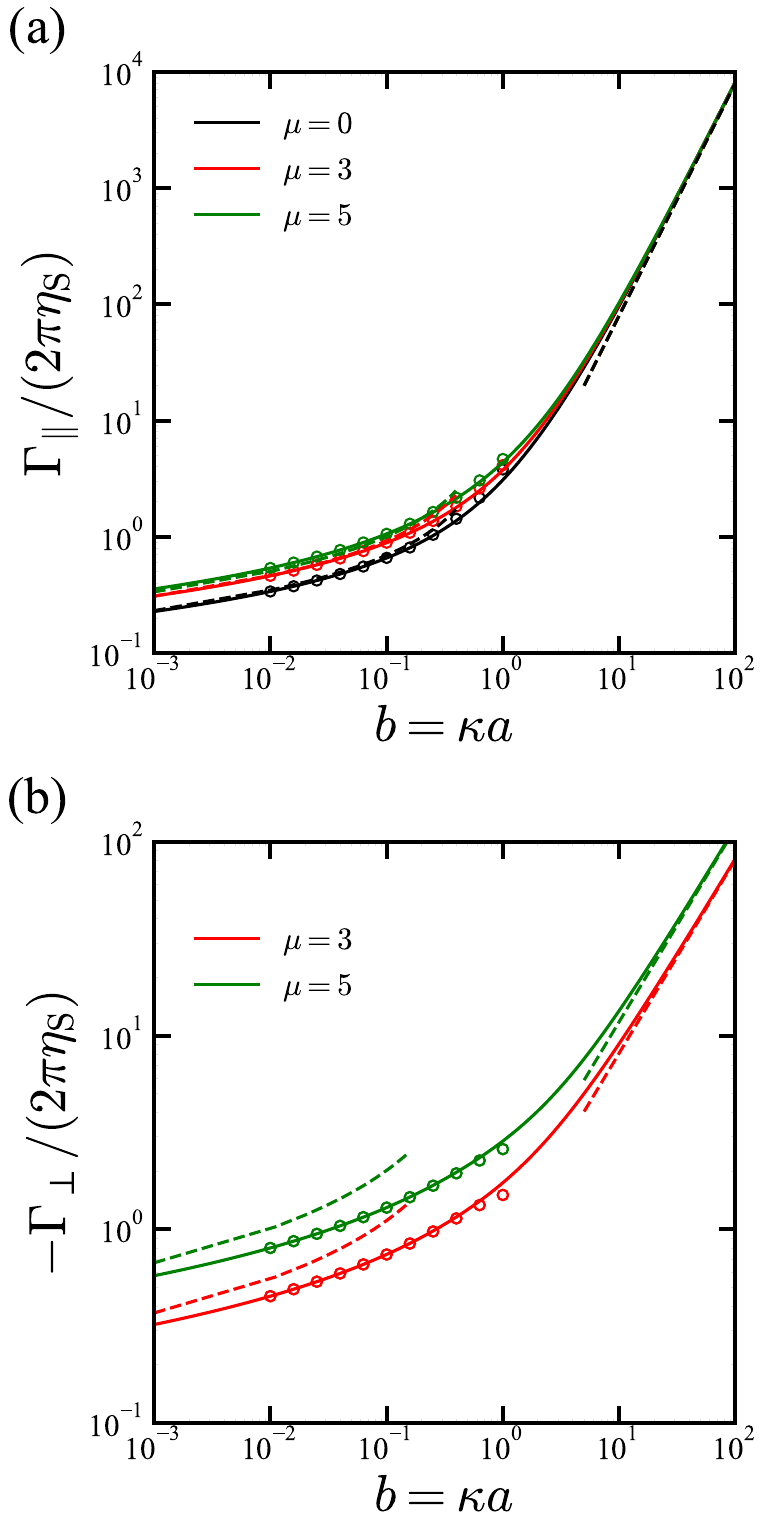} 
    \caption{Variation of (a) the resistance coefficient \( \Gamma_\| \) and (b) the lift coefficient \( -\Gamma_\perp \) with the rescaled disk radius \( b = \kappa a \) for different values of the odd-to-even viscosity ratio \( \mu = \eta_{\rm O}/\eta_{\rm S} \). The black, red, and green solid lines represent the full expressions for the resistance and lift coefficients with \( \mu = 0, 3, \) and \( 5 \), respectively, while keeping \( \xi = \kappa/\lambda = 1 \). 
    The asymptotic behaviors of \( \Gamma_\| \) and \( \Gamma_\perp \), as given by Eqs.~\eqref{eq:dragsmall}-\eqref{eq:liftlarge} for \( b\ll1 \) and \( b\gg1 \), are plotted with dashed lines. 
    The circle symbols represent the asymptotic expressions, as reported in Ref.~\onlinecite{hosaka2021nonreciprocal}.
    }
    \label{fig:resistance}
\end{figure}

Notably, the resistance force acting on the disk comprises two contributions: one from the resistance force due to the underlying bulk and the other from the thin fluid layer acting on the perimeter of the disk. Specifically,
\begin{equation}
    \bm{F}_\mathrm{R} = 
    \bm{F}_\mathrm{B} + \bm{F}_\mathrm{S} \, ,
\end{equation}
where
\begin{equation}
     \bm{F}_{\rm B} =\pi a^2 \bm{f}_{\rm B} \, ,
\end{equation}
and
\begin{equation}
 \bm{F}_\mathrm{S} = \int_0^{2\pi}  \left( \sigma_{rr} \, \hat{\bm{e}}_r + \sigma_{r\theta} \, \hat{\bm{e}}_\theta \right) \, a \, \mathrm{d}\theta \, .
\end{equation}

In polar coordinates, the shear stress tensor components are given by
\begin{subequations}
    \begin{align}
    \hspace{-0.15cm}
    \frac{\sigma_{rr}}{\eta_\mathrm{S}} &=
    \left( 3\beta+1 \right) \partial_r v_r 
    + \frac{3\beta-1}{r} \left( v_r + \partial_\theta v_\theta \right) \notag \\
    &\quad+ \mu \left( \partial_r v_\theta + 
    \frac{\partial_\theta v_r - v_\theta}{r}
    \right) , \\[3pt]
    \hspace{-0.15cm}   \frac{\sigma_{r\theta}}{\eta_\mathrm{S}} &=
    \partial_r v_\theta + 
    \frac{ \partial_\theta v_r - v_\theta }{r}
    + \mu \left( \frac{v_r+\partial_\theta v_\theta}{r} - \partial_r v_r \right).
\end{align}
\end{subequations}

We note that $\eta_\mathrm{D}/ \eta_\mathrm{S} = 3\beta$, so for $\beta = 1/3$ and $\mu = 0$, it follows that $\sigma_{ij} = 2\eta_\mathrm{S} E_{ij}$ (see Eq.~\eqref{eq:stress}), yielding the classical expressions of the stress tensor in polar coordinates.

The bulk contribution in terms of the force density can be expressed as
\begin{equation}
    \bm{F}_\mathrm{B} = -\pi \eta_\mathrm{S}
    V
    \left( b^2 \, \hat{\bm{e}}_x
    + \frac{3}{4} \left( \bm{B}-\bm{C} \right)
    \right) ,
\end{equation}
where $\bm{B}$ and~$\bm{C}$ are given by Eqs.~\eqref{eq:B} and~\eqref{eq:C}, respectively.

\begin{figure}
    \centering
    \includegraphics[width=0.85\linewidth]{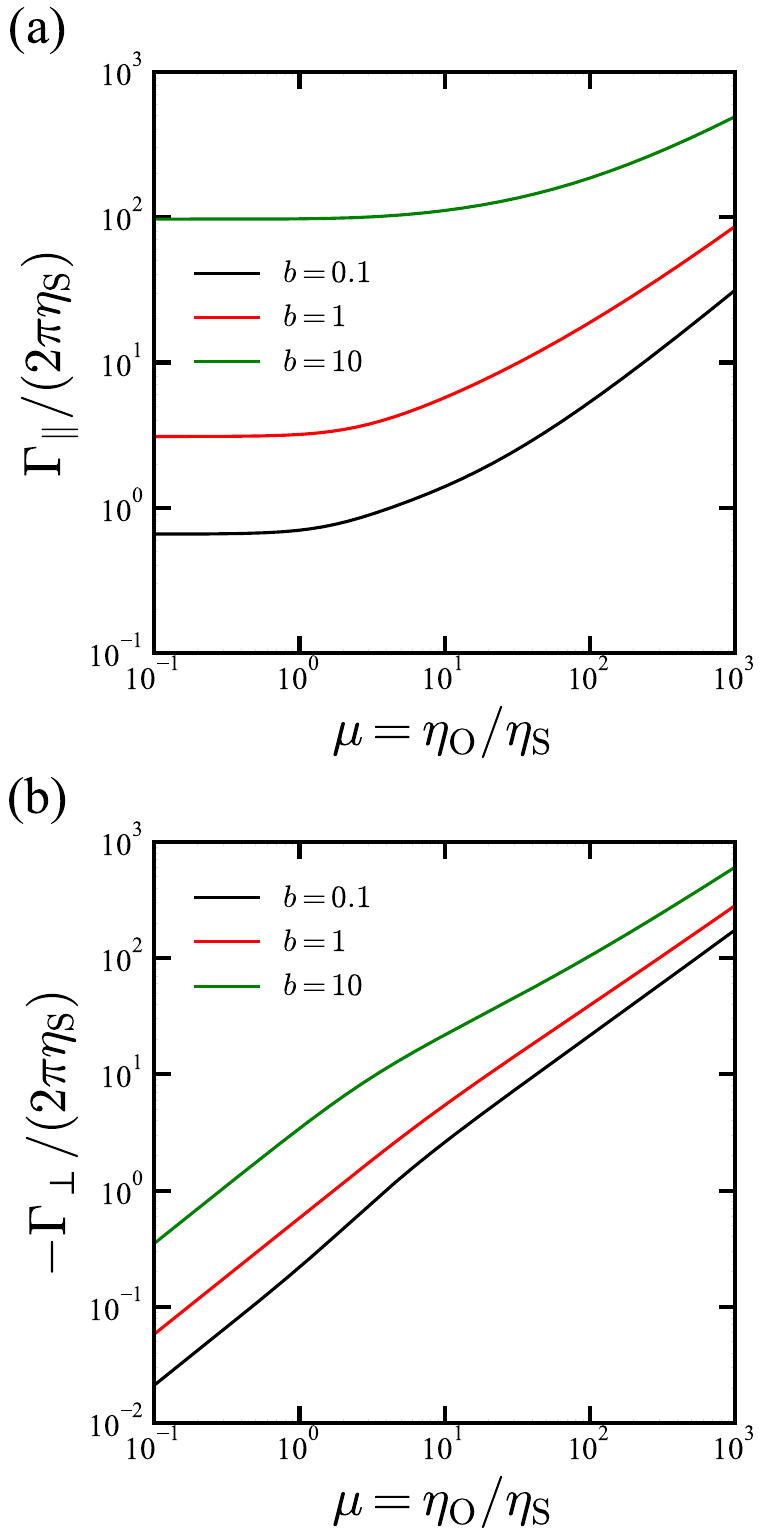} 
    \caption{Variation of (a) the resistance coefficient \( \Gamma_\| \) and (b) the lift coefficient \( -\Gamma_\perp \) as a function of \( \mu = \eta_{\rm O}/\eta_{\rm S} \) for different values of the rescaled disk radius \( b = \kappa a \). The black, red, and green solid lines correspond to \( b = 0.1, 1, \) and \( 10 \), respectively, while keeping \( \xi = \kappa/\lambda = 1 \).
    }
    \label{fig:resistancemu}
\end{figure}

For $b \ll 1$, we find that both $\Gamma_\parallel$ and $\Gamma_\perp$ scale inverse logarithmically with $b$ as
\begin{equation}
   \frac{\Gamma_\parallel}{2\pi\eta_\mathrm{S}} \simeq 
   -\frac{2}{\ln b} \frac{\left( 4\xi^2+1\right)\left( 4\xi^2+\mu^2\right)}{\left( 4\xi^2+1\right)^2 + 4\mu^2}
   \, ,
   \label{eq:dragsmall}
\end{equation}
for the resistance coefficient, and 
\begin{equation}
    \frac{\Gamma_\perp}{2\pi\eta_\mathrm{S}} \simeq
    \frac{4}{\ln b} \frac{\mu \left( 4\xi^2+\mu^2\right)}
    { \left( 4\xi^2+1\right)^2 + 4\mu^2}
    \, ,
    \label{eq:liftsmall}
\end{equation}
for the lift coefficient.

For $b \gg 1$, the resistance coefficient $\Gamma_\parallel$ scales as $b^2$ 
\begin{equation}
    \frac{\Gamma_\parallel}{2\pi\eta_\mathrm{S}} = \frac{4}{5} \, b^2 + \mathcal{O} (b)
    \, ,
    \label{eq:draglarge}
\end{equation}
which coincides with the result in Ref.~\onlinecite{barentin1999}.
Expression~\eqref{eq:liftsmall} becomes independent of the 2D viscosities, and depends only on $\eta_{\rm B}$, indicating that the lateral surface dynamics is governed by the supporting bulk fluid at large scales.
The lift coefficient $\Gamma_\perp$, on the other hand,  scales linearly with~$b$
\begin{equation}
    \frac{\Gamma_\perp}{2\pi\eta_\mathrm{S}} =
    -\frac{8}{25} \, b\mu 
    \left( 4\xi^2 + \mu^2\right)^\frac{1}{2}
    \left| \frac{A_+ - A_-}{\delta^\frac{1}{2} } \right| + \mathcal{O}(1)\, .
    \label{eq:liftlarge}
\end{equation}

In Fig.~\ref{fig:resistance}, we plot $\Gamma_\|$ and $-\Gamma_\perp$ as functions of the rescaled disk size $b$ for different values of the odd viscosity $\mu$.
Both resistance coefficients show only a weak dependence on disk size at small scales, while exhibiting a stronger size dependence with power-law behavior at larger scales.
The crossover between these two behaviors occurs approximately at $b \sim 1$.

The weak and logarithmic size dependence coincides with the resistance coefficient on a tracer moving in purely 2D fluids without~\cite{saffman1975, saffman1976, evans1988, barentin1999, diamant2009hydrodynamic} and with odd viscosity.\cite{hosaka2021hydrodynamic}
On the other hand, the quadratic scaling, $\Gamma_\|\sim b^2$, arises from momentum decay over distances beyond the screening lengths, $\kappa^{-1}$ and $\lambda^{-1}$, which is characteristic of a 2D fluid under confinement~\cite{evans1988}. The friction term, $-(\eta_{\rm B}/h)\bm{v}$, significantly suppresses momentum conservation in 2D, leading to a collapse of all $\Gamma_\|$ curves for $b \gg 1$, irrespective of the different values of the odd viscosity [Eq.~\eqref{eq:draglarge}].

In contrast, the friction term does not induce momentum decay along the transverse velocity $\boldsymbol{\epsilon} \cdot \bm{v}$. Consequently, even at sufficiently large scales ($b \gg 1$), momentum may be conserved in the underlying bulk fluid rather than within the 2D layer. This situation is typical for a 2D liquid or fluid membrane freely suspended in a 3D bulk medium without confinement.\cite{saffman1975, saffman1976, diamant2009hydrodynamic}. In this case, the mobility decays inversely proportional to the disk size,\cite{saffman1975, saffman1976, diamant2009hydrodynamic} implying the obtained linear scaling, $\Gamma_\perp \sim b$. Such a strong size dependence contrasts with the lift observed in a 2D liquid drop, where it becomes constant at large scales.\cite{hosaka2021hydrodynamic}

Our results closely match the asymptotic coefficients for smaller scales, as obtained from the boundary integral expression and the numerical evaluation of the Green's function.\cite{hosaka2021nonreciprocal}
We note a sign error in the lift coefficient reported in Ref.~\onlinecite{hosaka2021nonreciprocal}, as the derivation should have accounted for the generalized Lorentz reciprocal theorem in the presence of odd viscosity.\cite{hosaka2023lorentz}

Figure~\ref{fig:resistancemu} shows the resistance coefficients $\Gamma_\|$ and $\Gamma_\perp$ as functions of the odd-to-even viscosity ratio $\mu$ for various values of $b = \kappa a$. The resistance coefficient $\Gamma_\|$ remains nearly independent of $\mu$ for its small values, but varies monotonically for $\mu \gg 1$, as illustrated in Fig.~\ref{fig:resistancemu}(a).  
In contrast to the non-monotonic dependence of $\Gamma_\perp$ on $\mu$ observed for a 2D liquid drop,\cite{hosaka2021hydrodynamic} the lift on a compressible surface increases consistently with increasing $\mu$, as shown in Fig.~\ref{fig:resistancemu}(b).

The cylindrical symmetry about the \( z \)-direction implies the following form for the resistance tensor:
\begin{equation}
    \bm{\Gamma} = \Gamma_\| \mathds{1} - \Gamma_\perp \boldsymbol{\epsilon}\, ,
\end{equation}
which relates the disk velocity in any direction to the resistance force, \( \bm{F}_{\rm R} = -\bm{\Gamma} \cdot \bm{V} \). Since \( \Gamma_\| \) and \( \Gamma_\perp \) are even and odd functions of \( \mu \), respectively, we obtain the reciprocal relation \( \Gamma_{ij}(\mu) = \Gamma_{ji}(-\mu) \) by flipping the sign of the odd viscosity under index exchange. 

Such symmetry relations have been widely reported in systems with odd viscosity~\cite{hosaka2021nonreciprocal, hosaka2021hydrodynamic, lier2023lift, khain2024trading, everts2024dissipative, hosaka2024chirotactic} and are consistent with the Onsager-Casimir reciprocity~\cite{fruchart2023odd}, as required by the Lorentz reciprocal theorem for rigid objects of any shape in the presence of odd viscosity.\cite{hosaka2023lorentz}

\section{Concluding remarks} 
To date, experimental realizations of systems with odd viscosity have been found exclusively in 2D systems. A 2D compressible layer of odd viscosity, supported by a thin layer of conventional fluid, is therefore considered a standard model to study the odd response in such systems. Because odd viscosity couples the spatial degrees of freedom in a non-trivial way, exact solutions of problems involving odd viscosity remain very rare. Rather, the solutions are either carried out in the limit of small odd viscosity, or for objects that can be considered point particles. In this paper, we overcame these limitations and determined an exact analytical solution valid for a disk of arbitrary size. The only assumption required is that the supportive layer is thin and can therefore be described with the lubrication approximation. We could demonstrate that while breaking the Onsager reciprocity, the resulting resistance matrix  satisfies the Onsager-Casimir reciprocity to any order in odd viscosity. We expect that the general solution will facilitate direct experimental verification of odd properties in systems where rotating particles (microorganisms, rotary motor proteins or magnetic particles) are suspended in the proximity of an interface. At the same time, our work paves the way to further exact analytical solutions of hydrodynamic problems involving multiple characteristic length scales. 

\begin{acknowledgments}
A.V.\ acknowledges support from the Slovenian
Research Agency (Grant No. P1-0099).
\end{acknowledgments}

\section*{Data Availability Statement}
The data that support the findings of this study are available from the corresponding author upon reasonable request. 

\section*{References}


\begin{thebibliography}{61}%
\makeatletter
\providecommand \@ifxundefined [1]{%
 \@ifx{#1\undefined}
}%
\providecommand \@ifnum [1]{%
 \ifnum #1\expandafter \@firstoftwo
 \else \expandafter \@secondoftwo
 \fi
}%
\providecommand \@ifx [1]{%
 \ifx #1\expandafter \@firstoftwo
 \else \expandafter \@secondoftwo
 \fi
}%
\providecommand \natexlab [1]{#1}%
\providecommand \enquote  [1]{``#1''}%
\providecommand \bibnamefont  [1]{#1}%
\providecommand \bibfnamefont [1]{#1}%
\providecommand \citenamefont [1]{#1}%
\providecommand \href@noop [0]{\@secondoftwo}%
\providecommand \href [0]{\begingroup \@sanitize@url \@href}%
\providecommand \@href[1]{\@@startlink{#1}\@@href}%
\providecommand \@@href[1]{\endgroup#1\@@endlink}%
\providecommand \@sanitize@url [0]{\catcode `\\12\catcode `\$12\catcode
  `\&12\catcode `\#12\catcode `\^12\catcode `\_12\catcode `\%12\relax}%
\providecommand \@@startlink[1]{}%
\providecommand \@@endlink[0]{}%
\providecommand \url  [0]{\begingroup\@sanitize@url \@url }%
\providecommand \@url [1]{\endgroup\@href {#1}{\urlprefix }}%
\providecommand \urlprefix  [0]{URL }%
\providecommand \Eprint [0]{\href }%
\providecommand \doibase [0]{https://doi.org/}%
\providecommand \selectlanguage [0]{\@gobble}%
\providecommand \bibinfo  [0]{\@secondoftwo}%
\providecommand \bibfield  [0]{\@secondoftwo}%
\providecommand \translation [1]{[#1]}%
\providecommand \BibitemOpen [0]{}%
\providecommand \bibitemStop [0]{}%
\providecommand \bibitemNoStop [0]{.\EOS\space}%
\providecommand \EOS [0]{\spacefactor3000\relax}%
\providecommand \BibitemShut  [1]{\csname bibitem#1\endcsname}%
\let\auto@bib@innerbib\@empty
\bibitem [{\citenamefont {Marchetti}\ \emph {et~al.}(2013)\citenamefont
  {Marchetti}, \citenamefont {Joanny}, \citenamefont {Ramaswamy}, \citenamefont
  {Liverpool}, \citenamefont {Prost}, \citenamefont {Rao},\ and\ \citenamefont
  {Simha}}]{marchetti2013hydrodynamics}%
  \BibitemOpen
  \bibfield  {author} {\bibinfo {author} {\bibfnamefont {M.~C.}\ \bibnamefont
  {Marchetti}}, \bibinfo {author} {\bibfnamefont {J.-F.}\ \bibnamefont
  {Joanny}}, \bibinfo {author} {\bibfnamefont {S.}~\bibnamefont {Ramaswamy}},
  \bibinfo {author} {\bibfnamefont {T.~B.}\ \bibnamefont {Liverpool}}, \bibinfo
  {author} {\bibfnamefont {J.}~\bibnamefont {Prost}}, \bibinfo {author}
  {\bibfnamefont {M.}~\bibnamefont {Rao}},\ and\ \bibinfo {author}
  {\bibfnamefont {R.~A.}\ \bibnamefont {Simha}},\ }\bibfield  {title} {\enquote
  {\bibinfo {title} {Hydrodynamics of soft active matter},}\ }\href
  {https://doi.org/https://doi.org/10.1103/RevModPhys.85.1143} {\bibfield
  {journal} {\bibinfo  {journal} {Rev. Mod. Phys.}\ }\textbf {\bibinfo {volume}
  {85}},\ \bibinfo {pages} {1143} (\bibinfo {year} {2013})}\BibitemShut
  {NoStop}%
\bibitem [{\citenamefont {Bechinger}\ \emph {et~al.}(2016)\citenamefont
  {Bechinger}, \citenamefont {Di~Leonardo}, \citenamefont {L{\"o}wen},
  \citenamefont {Reichhardt}, \citenamefont {Volpe},\ and\ \citenamefont
  {Volpe}}]{bechinger2016active}%
  \BibitemOpen
  \bibfield  {author} {\bibinfo {author} {\bibfnamefont {C.}~\bibnamefont
  {Bechinger}}, \bibinfo {author} {\bibfnamefont {R.}~\bibnamefont
  {Di~Leonardo}}, \bibinfo {author} {\bibfnamefont {H.}~\bibnamefont
  {L{\"o}wen}}, \bibinfo {author} {\bibfnamefont {C.}~\bibnamefont
  {Reichhardt}}, \bibinfo {author} {\bibfnamefont {G.}~\bibnamefont {Volpe}},\
  and\ \bibinfo {author} {\bibfnamefont {G.}~\bibnamefont {Volpe}},\ }\bibfield
   {title} {\enquote {\bibinfo {title} {Active particles in complex and crowded
  environments},}\ }\href
  {https://doi.org/https://doi.org/10.1103/RevModPhys.88.045006} {\bibfield
  {journal} {\bibinfo  {journal} {Rev. Mod. Phys.}\ }\textbf {\bibinfo {volume}
  {88}},\ \bibinfo {pages} {045006} (\bibinfo {year} {2016})}\BibitemShut
  {NoStop}%
\bibitem [{\citenamefont {Cates}\ and\ \citenamefont
  {Tailleur}(2015)}]{cates2015motility}%
  \BibitemOpen
  \bibfield  {author} {\bibinfo {author} {\bibfnamefont {M.~E.}\ \bibnamefont
  {Cates}}\ and\ \bibinfo {author} {\bibfnamefont {J.}~\bibnamefont
  {Tailleur}},\ }\bibfield  {title} {\enquote {\bibinfo {title}
  {Motility-induced phase separation},}\ }\href
  {https://doi.org/10.1146/annurev-conmatphys-031214-014710} {\bibfield
  {journal} {\bibinfo  {journal} {Annu. Rev. Condens. Matter Phys.}\ }\textbf
  {\bibinfo {volume} {6}},\ \bibinfo {pages} {219--244} (\bibinfo {year}
  {2015})}\BibitemShut {NoStop}%
\bibitem [{\citenamefont {Digregorio}\ \emph {et~al.}(2018)\citenamefont
  {Digregorio}, \citenamefont {Levis}, \citenamefont {Suma}, \citenamefont
  {Cugliandolo}, \citenamefont {Gonnella},\ and\ \citenamefont
  {Pagonabarraga}}]{digregorio2018full}%
  \BibitemOpen
  \bibfield  {author} {\bibinfo {author} {\bibfnamefont {P.}~\bibnamefont
  {Digregorio}}, \bibinfo {author} {\bibfnamefont {D.}~\bibnamefont {Levis}},
  \bibinfo {author} {\bibfnamefont {A.}~\bibnamefont {Suma}}, \bibinfo {author}
  {\bibfnamefont {L.~F.}\ \bibnamefont {Cugliandolo}}, \bibinfo {author}
  {\bibfnamefont {G.}~\bibnamefont {Gonnella}},\ and\ \bibinfo {author}
  {\bibfnamefont {I.}~\bibnamefont {Pagonabarraga}},\ }\bibfield  {title}
  {\enquote {\bibinfo {title} {Full phase diagram of active {Brownian} disks:
  from melting to motility-induced phase separation},}\ }\href
  {https://doi.org/https://doi.org/10.1103/PhysRevLett.121.098003} {\bibfield
  {journal} {\bibinfo  {journal} {Phys. Rev. Lett.}\ }\textbf {\bibinfo
  {volume} {121}},\ \bibinfo {pages} {098003} (\bibinfo {year}
  {2018})}\BibitemShut {NoStop}%
\bibitem [{\citenamefont {Paoluzzi}, \citenamefont {Levis},\ and\ \citenamefont
  {Pagonabarraga}(2022)}]{paoluzzi2022motility}%
  \BibitemOpen
  \bibfield  {author} {\bibinfo {author} {\bibfnamefont {M.}~\bibnamefont
  {Paoluzzi}}, \bibinfo {author} {\bibfnamefont {D.}~\bibnamefont {Levis}},\
  and\ \bibinfo {author} {\bibfnamefont {I.}~\bibnamefont {Pagonabarraga}},\
  }\bibfield  {title} {\enquote {\bibinfo {title} {From motility-induced
  phase-separation to glassiness in dense active matter},}\ }\href
  {https://doi.org/https://doi.org/10.1038/s42005-022-00886-3} {\bibfield
  {journal} {\bibinfo  {journal} {Commun. Phys.}\ }\textbf {\bibinfo {volume}
  {5}},\ \bibinfo {pages} {111} (\bibinfo {year} {2022})}\BibitemShut {NoStop}%
\bibitem [{\citenamefont {Sharan}\ \emph {et~al.}(2023)\citenamefont {Sharan},
  \citenamefont {Daddi-Moussa-Ider}, \citenamefont {Agudo-Canalejo},
  \citenamefont {Golestanian},\ and\ \citenamefont
  {Simmchen}}]{sharan2023pair}%
  \BibitemOpen
  \bibfield  {author} {\bibinfo {author} {\bibfnamefont {P.}~\bibnamefont
  {Sharan}}, \bibinfo {author} {\bibfnamefont {A.}~\bibnamefont
  {Daddi-Moussa-Ider}}, \bibinfo {author} {\bibfnamefont {J.}~\bibnamefont
  {Agudo-Canalejo}}, \bibinfo {author} {\bibfnamefont {R.}~\bibnamefont
  {Golestanian}},\ and\ \bibinfo {author} {\bibfnamefont {J.}~\bibnamefont
  {Simmchen}},\ }\bibfield  {title} {\enquote {\bibinfo {title} {Pair
  interaction between two catalytically active colloids},}\ }\href
  {https://doi.org/10.1002/smll.202300817} {\bibfield  {journal} {\bibinfo
  {journal} {Small}\ }\textbf {\bibinfo {volume} {19}},\ \bibinfo {pages}
  {2300817} (\bibinfo {year} {2023})}\BibitemShut {NoStop}%
\bibitem [{\citenamefont {Denk}\ and\ \citenamefont
  {Frey}(2020)}]{denk2020pattern}%
  \BibitemOpen
  \bibfield  {author} {\bibinfo {author} {\bibfnamefont {J.}~\bibnamefont
  {Denk}}\ and\ \bibinfo {author} {\bibfnamefont {E.}~\bibnamefont {Frey}},\
  }\bibfield  {title} {\enquote {\bibinfo {title} {Pattern-induced local
  symmetry breaking in active-matter systems},}\ }\href
  {https://doi.org/https://doi.org/10.1073/pnas.2010302117} {\bibfield
  {journal} {\bibinfo  {journal} {Proc. Natl. Acad. Sci.}\ }\textbf {\bibinfo
  {volume} {117}},\ \bibinfo {pages} {31623--31630} (\bibinfo {year}
  {2020})}\BibitemShut {NoStop}%
\bibitem [{\citenamefont {Tjhung}, \citenamefont {Marenduzzo},\ and\
  \citenamefont {Cates}(2012)}]{tjhung2012spontaneous}%
  \BibitemOpen
  \bibfield  {author} {\bibinfo {author} {\bibfnamefont {E.}~\bibnamefont
  {Tjhung}}, \bibinfo {author} {\bibfnamefont {D.}~\bibnamefont {Marenduzzo}},\
  and\ \bibinfo {author} {\bibfnamefont {M.~E.}\ \bibnamefont {Cates}},\
  }\bibfield  {title} {\enquote {\bibinfo {title} {Spontaneous symmetry
  breaking in active droplets provides a generic route to motility},}\ }\href
  {https://doi.org/https://doi.org/10.1073/pnas.1200843109} {\bibfield
  {journal} {\bibinfo  {journal} {Proc. Natl. Acad. Sci.}\ }\textbf {\bibinfo
  {volume} {109}},\ \bibinfo {pages} {12381--12386} (\bibinfo {year}
  {2012})}\BibitemShut {NoStop}%
\bibitem [{\citenamefont {Shankar}\ \emph {et~al.}(2022)\citenamefont
  {Shankar}, \citenamefont {Souslov}, \citenamefont {Bowick}, \citenamefont
  {Marchetti},\ and\ \citenamefont {Vitelli}}]{shankar2022topological}%
  \BibitemOpen
  \bibfield  {author} {\bibinfo {author} {\bibfnamefont {S.}~\bibnamefont
  {Shankar}}, \bibinfo {author} {\bibfnamefont {A.}~\bibnamefont {Souslov}},
  \bibinfo {author} {\bibfnamefont {M.~J.}\ \bibnamefont {Bowick}}, \bibinfo
  {author} {\bibfnamefont {M.~C.}\ \bibnamefont {Marchetti}},\ and\ \bibinfo
  {author} {\bibfnamefont {V.}~\bibnamefont {Vitelli}},\ }\bibfield  {title}
  {\enquote {\bibinfo {title} {Topological active matter},}\ }\href
  {https://doi.org/10.1038/s42254-022-00445-3} {\bibfield  {journal} {\bibinfo
  {journal} {Nat. Rev. Phys.}\ }\textbf {\bibinfo {volume} {4}},\ \bibinfo
  {pages} {380--398} (\bibinfo {year} {2022})}\BibitemShut {NoStop}%
\bibitem [{\citenamefont {F{\"u}rthauer}\ \emph {et~al.}(2012)\citenamefont
  {F{\"u}rthauer}, \citenamefont {Strempel}, \citenamefont {Grill},\ and\
  \citenamefont {J{\"u}licher}}]{furthauer2012active}%
  \BibitemOpen
  \bibfield  {author} {\bibinfo {author} {\bibfnamefont {S.}~\bibnamefont
  {F{\"u}rthauer}}, \bibinfo {author} {\bibfnamefont {M.}~\bibnamefont
  {Strempel}}, \bibinfo {author} {\bibfnamefont {S.~W.}\ \bibnamefont
  {Grill}},\ and\ \bibinfo {author} {\bibfnamefont {F.}~\bibnamefont
  {J{\"u}licher}},\ }\bibfield  {title} {\enquote {\bibinfo {title} {Active
  chiral fluids},}\ }\href {https://doi.org/10.1140/epje/i2012-12089-6}
  {\bibfield  {journal} {\bibinfo  {journal} {Eur. Phys. J. E}\ }\textbf
  {\bibinfo {volume} {35}},\ \bibinfo {pages} {89} (\bibinfo {year}
  {2012})}\BibitemShut {NoStop}%
\bibitem [{\citenamefont {Liebchen}\ and\ \citenamefont
  {Levis}(2022)}]{liebchen2022chiral}%
  \BibitemOpen
  \bibfield  {author} {\bibinfo {author} {\bibfnamefont {B.}~\bibnamefont
  {Liebchen}}\ and\ \bibinfo {author} {\bibfnamefont {D.}~\bibnamefont
  {Levis}},\ }\bibfield  {title} {\enquote {\bibinfo {title} {Chiral active
  matter},}\ }\href {https://doi.org/10.1209/0295-5075/ac8f69} {\bibfield
  {journal} {\bibinfo  {journal} {Europhys. Lett.}\ }\textbf {\bibinfo {volume}
  {139}},\ \bibinfo {pages} {67001} (\bibinfo {year} {2022})}\BibitemShut
  {NoStop}%
\bibitem [{\citenamefont {Hosaka}\ and\ \citenamefont
  {Komura}(2022)}]{hosaka2022nonequilibrium}%
  \BibitemOpen
  \bibfield  {author} {\bibinfo {author} {\bibfnamefont {Y.}~\bibnamefont
  {Hosaka}}\ and\ \bibinfo {author} {\bibfnamefont {S.}~\bibnamefont
  {Komura}},\ }\bibfield  {title} {\enquote {\bibinfo {title} {Nonequilibrium
  transport induced by biological nanomachines},}\ }\href
  {https://doi.org/https://doi.org/10.1142/S1793048022310026} {\bibfield
  {journal} {\bibinfo  {journal} {Biophys. Rev. Lett.}\ }\textbf {\bibinfo
  {volume} {17}},\ \bibinfo {pages} {51} (\bibinfo {year} {2022})}\BibitemShut
  {NoStop}%
\bibitem [{\citenamefont {Fruchart}, \citenamefont {Scheibner},\ and\
  \citenamefont {Vitelli}(2023)}]{fruchart2023odd}%
  \BibitemOpen
  \bibfield  {author} {\bibinfo {author} {\bibfnamefont {M.}~\bibnamefont
  {Fruchart}}, \bibinfo {author} {\bibfnamefont {C.}~\bibnamefont
  {Scheibner}},\ and\ \bibinfo {author} {\bibfnamefont {V.}~\bibnamefont
  {Vitelli}},\ }\bibfield  {title} {\enquote {\bibinfo {title} {Odd viscosity
  and odd elasticity},}\ }\href
  {https://doi.org/10.1146/annurev-conmatphys-040821-125506} {\bibfield
  {journal} {\bibinfo  {journal} {Annu. Rev. Condens. Matter Phys.}\ }\textbf
  {\bibinfo {volume} {14}},\ \bibinfo {pages} {471--510} (\bibinfo {year}
  {2023})}\BibitemShut {NoStop}%
\bibitem [{\citenamefont {Mecke}\ \emph {et~al.}(2024)\citenamefont {Mecke},
  \citenamefont {Nketsiah}, \citenamefont {Li},\ and\ \citenamefont
  {Gao}}]{mecke2024emergent}%
  \BibitemOpen
  \bibfield  {author} {\bibinfo {author} {\bibfnamefont {J.}~\bibnamefont
  {Mecke}}, \bibinfo {author} {\bibfnamefont {J.~O.}\ \bibnamefont {Nketsiah}},
  \bibinfo {author} {\bibfnamefont {R.}~\bibnamefont {Li}},\ and\ \bibinfo
  {author} {\bibfnamefont {Y.}~\bibnamefont {Gao}},\ }\bibfield  {title}
  {\enquote {\bibinfo {title} {Emergent phenomena in chiral active matter},}\
  }\href {https://doi.org/https://doi.org/10.1360/nso/20230086} {\bibfield
  {journal} {\bibinfo  {journal} {National Science Open}\ }\textbf {\bibinfo
  {volume} {3}},\ \bibinfo {pages} {20230086} (\bibinfo {year}
  {2024})}\BibitemShut {NoStop}%
\bibitem [{\citenamefont {Caprini}, \citenamefont {Liebchen},\ and\
  \citenamefont {L{\"o}wen}(2024)}]{caprini2024self}%
  \BibitemOpen
  \bibfield  {author} {\bibinfo {author} {\bibfnamefont {L.}~\bibnamefont
  {Caprini}}, \bibinfo {author} {\bibfnamefont {B.}~\bibnamefont {Liebchen}},\
  and\ \bibinfo {author} {\bibfnamefont {H.}~\bibnamefont {L{\"o}wen}},\
  }\bibfield  {title} {\enquote {\bibinfo {title} {Self-reverting vortices in
  chiral active matter},}\ }\href
  {https://doi.org/https://doi.org/10.1038/s42005-024-01637-2} {\bibfield
  {journal} {\bibinfo  {journal} {Commun. Physics}\ }\textbf {\bibinfo {volume}
  {7}},\ \bibinfo {pages} {153} (\bibinfo {year} {2024})}\BibitemShut {NoStop}%
\bibitem [{\citenamefont {Soni}\ \emph {et~al.}(2019)\citenamefont {Soni},
  \citenamefont {Bililign}, \citenamefont {Magkiriadou}, \citenamefont
  {Sacanna}, \citenamefont {Bartolo}, \citenamefont {Shelley},\ and\
  \citenamefont {Irvine}}]{soni2019odd}%
  \BibitemOpen
  \bibfield  {author} {\bibinfo {author} {\bibfnamefont {V.}~\bibnamefont
  {Soni}}, \bibinfo {author} {\bibfnamefont {E.~S.}\ \bibnamefont {Bililign}},
  \bibinfo {author} {\bibfnamefont {S.}~\bibnamefont {Magkiriadou}}, \bibinfo
  {author} {\bibfnamefont {S.}~\bibnamefont {Sacanna}}, \bibinfo {author}
  {\bibfnamefont {D.}~\bibnamefont {Bartolo}}, \bibinfo {author} {\bibfnamefont
  {M.~J.}\ \bibnamefont {Shelley}},\ and\ \bibinfo {author} {\bibfnamefont
  {W.~T.~M.}\ \bibnamefont {Irvine}},\ }\bibfield  {title} {\enquote {\bibinfo
  {title} {The odd free surface flows of a colloidal chiral fluid},}\ }\href
  {https://doi.org/10.1038/s41567-019-0603-8} {\bibfield  {journal} {\bibinfo
  {journal} {Nat. Phys.}\ }\textbf {\bibinfo {volume} {15}},\ \bibinfo {pages}
  {1188--1194} (\bibinfo {year} {2019})}\BibitemShut {NoStop}%
\bibitem [{\citenamefont {L{\'o}pez-Casta{\~n}o}\ \emph
  {et~al.}(2022)\citenamefont {L{\'o}pez-Casta{\~n}o}, \citenamefont
  {M{\'a}rquez~Seco}, \citenamefont {M{\'a}rquez~Seco}, \citenamefont
  {Rodr{\'\i}guez-Rivas},\ and\ \citenamefont {Reyes}}]{lopez2022chirality}%
  \BibitemOpen
  \bibfield  {author} {\bibinfo {author} {\bibfnamefont {M.~A.}\ \bibnamefont
  {L{\'o}pez-Casta{\~n}o}}, \bibinfo {author} {\bibfnamefont {A.}~\bibnamefont
  {M{\'a}rquez~Seco}}, \bibinfo {author} {\bibfnamefont {A.}~\bibnamefont
  {M{\'a}rquez~Seco}}, \bibinfo {author} {\bibfnamefont {{\'A}.}~\bibnamefont
  {Rodr{\'\i}guez-Rivas}},\ and\ \bibinfo {author} {\bibfnamefont {F.~V.}\
  \bibnamefont {Reyes}},\ }\bibfield  {title} {\enquote {\bibinfo {title}
  {Chirality transitions in a system of active flat spinners},}\ }\href
  {https://doi.org/https://doi.org/10.1103/PhysRevResearch.4.033230} {\bibfield
   {journal} {\bibinfo  {journal} {Phys. Rev. Research}\ }\textbf {\bibinfo
  {volume} {4}},\ \bibinfo {pages} {033230} (\bibinfo {year}
  {2022})}\BibitemShut {NoStop}%
\bibitem [{\citenamefont {Zhao}\ \emph {et~al.}(2021)\citenamefont {Zhao},
  \citenamefont {Wang}, \citenamefont {Komura}, \citenamefont {Yang},
  \citenamefont {Ye},\ and\ \citenamefont {Seto}}]{zhao2021}%
  \BibitemOpen
  \bibfield  {author} {\bibinfo {author} {\bibfnamefont {Z.}~\bibnamefont
  {Zhao}}, \bibinfo {author} {\bibfnamefont {B.}~\bibnamefont {Wang}}, \bibinfo
  {author} {\bibfnamefont {S.}~\bibnamefont {Komura}}, \bibinfo {author}
  {\bibfnamefont {M.}~\bibnamefont {Yang}}, \bibinfo {author} {\bibfnamefont
  {F.}~\bibnamefont {Ye}},\ and\ \bibinfo {author} {\bibfnamefont
  {R.}~\bibnamefont {Seto}},\ }\bibfield  {title} {\enquote {\bibinfo {title}
  {Emergent stripes of active rotors in shear flows},}\ }\href
  {https://doi.org/https://doi.org/10.1103/PhysRevResearch.3.043229} {\bibfield
   {journal} {\bibinfo  {journal} {Phys. Rev. Research}\ }\textbf {\bibinfo
  {volume} {3}},\ \bibinfo {pages} {043229} (\bibinfo {year}
  {2021})}\BibitemShut {NoStop}%
\bibitem [{\citenamefont {Mecke}\ \emph {et~al.}(2023)\citenamefont {Mecke},
  \citenamefont {Gao}, \citenamefont {Ram{\'\i}rez~Medina}, \citenamefont
  {Aarts}, \citenamefont {Gompper},\ and\ \citenamefont
  {Ripoll}}]{mecke2023simultaneous}%
  \BibitemOpen
  \bibfield  {author} {\bibinfo {author} {\bibfnamefont {J.}~\bibnamefont
  {Mecke}}, \bibinfo {author} {\bibfnamefont {Y.}~\bibnamefont {Gao}}, \bibinfo
  {author} {\bibfnamefont {C.~A.}\ \bibnamefont {Ram{\'\i}rez~Medina}},
  \bibinfo {author} {\bibfnamefont {D.~G.}\ \bibnamefont {Aarts}}, \bibinfo
  {author} {\bibfnamefont {G.}~\bibnamefont {Gompper}},\ and\ \bibinfo {author}
  {\bibfnamefont {M.}~\bibnamefont {Ripoll}},\ }\bibfield  {title} {\enquote
  {\bibinfo {title} {Simultaneous emergence of active turbulence and odd
  viscosity in a colloidal chiral active system},}\ }\href
  {https://doi.org/10.1038/s42005-023-01442-3} {\bibfield  {journal} {\bibinfo
  {journal} {Commun. Phys.}\ }\textbf {\bibinfo {volume} {6}},\ \bibinfo
  {pages} {324} (\bibinfo {year} {2023})}\BibitemShut {NoStop}%
\bibitem [{\citenamefont {Chen}\ \emph
  {et~al.}(2024{\natexlab{a}})\citenamefont {Chen}, \citenamefont {Weady},
  \citenamefont {Atis}, \citenamefont {Matsuzawa}, \citenamefont {Shelley},\
  and\ \citenamefont {Irvine}}]{chen2024self}%
  \BibitemOpen
  \bibfield  {author} {\bibinfo {author} {\bibfnamefont {P.}~\bibnamefont
  {Chen}}, \bibinfo {author} {\bibfnamefont {S.}~\bibnamefont {Weady}},
  \bibinfo {author} {\bibfnamefont {S.}~\bibnamefont {Atis}}, \bibinfo {author}
  {\bibfnamefont {T.}~\bibnamefont {Matsuzawa}}, \bibinfo {author}
  {\bibfnamefont {M.~J.}\ \bibnamefont {Shelley}},\ and\ \bibinfo {author}
  {\bibfnamefont {W.~T.}\ \bibnamefont {Irvine}},\ }\bibfield  {title}
  {\enquote {\bibinfo {title} {Self-propulsion, flocking and chiral active
  phases from particles spinning at intermediate {Reynolds} numbers},}\ }\href
  {https://doi.org/https://doi.org/10.1038/s41567-024-02651-5} {\bibfield
  {journal} {\bibinfo  {journal} {Nat. Phys.}\ ,\ \bibinfo {pages} {1--9}}
  (\bibinfo {year} {2024}{\natexlab{a}})}\BibitemShut {NoStop}%
\bibitem [{\citenamefont {Zhao}\ \emph {et~al.}(2022)\citenamefont {Zhao},
  \citenamefont {Yang}, \citenamefont {Komura},\ and\ \citenamefont
  {Seto}}]{zhao2022odd}%
  \BibitemOpen
  \bibfield  {author} {\bibinfo {author} {\bibfnamefont {Z.}~\bibnamefont
  {Zhao}}, \bibinfo {author} {\bibfnamefont {M.}~\bibnamefont {Yang}}, \bibinfo
  {author} {\bibfnamefont {S.}~\bibnamefont {Komura}},\ and\ \bibinfo {author}
  {\bibfnamefont {R.}~\bibnamefont {Seto}},\ }\bibfield  {title} {\enquote
  {\bibinfo {title} {Odd viscosity in chiral passive suspensions},}\ }\href
  {https://doi.org/https://doi.org/10.3389/fphy.2022.951465} {\bibfield
  {journal} {\bibinfo  {journal} {Front. Phys.}\ }\textbf {\bibinfo {volume}
  {10}},\ \bibinfo {pages} {951465} (\bibinfo {year} {2022})}\BibitemShut
  {NoStop}%
\bibitem [{\citenamefont {Markovich}\ and\ \citenamefont
  {Lubensky}(2021)}]{markovich2021}%
  \BibitemOpen
  \bibfield  {author} {\bibinfo {author} {\bibfnamefont {T.}~\bibnamefont
  {Markovich}}\ and\ \bibinfo {author} {\bibfnamefont {T.~C.}\ \bibnamefont
  {Lubensky}},\ }\bibfield  {title} {\enquote {\bibinfo {title} {Odd viscosity
  in active matter: Microscopic origin and 3{D} effects},}\ }\href
  {https://doi.org/10.1103/PhysRevLett.127.048001} {\bibfield  {journal}
  {\bibinfo  {journal} {Phys. Rev. Lett.}\ }\textbf {\bibinfo {volume} {127}},\
  \bibinfo {pages} {048001} (\bibinfo {year} {2021})}\BibitemShut {NoStop}%
\bibitem [{\citenamefont {Avron}(1998)}]{avron1998}%
  \BibitemOpen
  \bibfield  {author} {\bibinfo {author} {\bibfnamefont {J.~E.}\ \bibnamefont
  {Avron}},\ }\bibfield  {title} {\enquote {\bibinfo {title} {Odd viscosity},}\
  }\href {https://doi.org/10.1023/A:1023084404080} {\bibfield  {journal}
  {\bibinfo  {journal} {J. Stat. Phys.}\ }\textbf {\bibinfo {volume} {92}},\
  \bibinfo {pages} {543--557} (\bibinfo {year} {1998})}\BibitemShut {NoStop}%
\bibitem [{\citenamefont {Banerjee}\ \emph {et~al.}(2017)\citenamefont
  {Banerjee}, \citenamefont {Souslov}, \citenamefont {Abanov},\ and\
  \citenamefont {Vitelli}}]{banerjee2017}%
  \BibitemOpen
  \bibfield  {author} {\bibinfo {author} {\bibfnamefont {D.}~\bibnamefont
  {Banerjee}}, \bibinfo {author} {\bibfnamefont {A.}~\bibnamefont {Souslov}},
  \bibinfo {author} {\bibfnamefont {A.~G.}\ \bibnamefont {Abanov}},\ and\
  \bibinfo {author} {\bibfnamefont {V.}~\bibnamefont {Vitelli}},\ }\bibfield
  {title} {\enquote {\bibinfo {title} {Odd viscosity in chiral active
  fluids},}\ }\href {https://doi.org/10.1038/s41467-017-01378-7} {\bibfield
  {journal} {\bibinfo  {journal} {Nat. Commun.}\ }\textbf {\bibinfo {volume}
  {8}},\ \bibinfo {pages} {1573} (\bibinfo {year} {2017})}\BibitemShut
  {NoStop}%
\bibitem [{\citenamefont {Souslov}\ \emph {et~al.}(2019)\citenamefont
  {Souslov}, \citenamefont {Dasbiswas}, \citenamefont {Fruchart}, \citenamefont
  {Vaikuntanathan},\ and\ \citenamefont {Vitelli}}]{souslov2019topological}%
  \BibitemOpen
  \bibfield  {author} {\bibinfo {author} {\bibfnamefont {A.}~\bibnamefont
  {Souslov}}, \bibinfo {author} {\bibfnamefont {K.}~\bibnamefont {Dasbiswas}},
  \bibinfo {author} {\bibfnamefont {M.}~\bibnamefont {Fruchart}}, \bibinfo
  {author} {\bibfnamefont {S.}~\bibnamefont {Vaikuntanathan}},\ and\ \bibinfo
  {author} {\bibfnamefont {V.}~\bibnamefont {Vitelli}},\ }\bibfield  {title}
  {\enquote {\bibinfo {title} {Topological waves in fluids with odd
  viscosity},}\ }\href {https://doi.org/10.1103/PhysRevLett.122.128001}
  {\bibfield  {journal} {\bibinfo  {journal} {Phys. Rev. Lett.}\ }\textbf
  {\bibinfo {volume} {122}},\ \bibinfo {pages} {128001} (\bibinfo {year}
  {2019})}\BibitemShut {NoStop}%
\bibitem [{\citenamefont {de~Wit}\ \emph {et~al.}(2024)\citenamefont {de~Wit},
  \citenamefont {Fruchart}, \citenamefont {Khain}, \citenamefont {Toschi},\
  and\ \citenamefont {Vitelli}}]{de2024pattern}%
  \BibitemOpen
  \bibfield  {author} {\bibinfo {author} {\bibfnamefont {X.~M.}\ \bibnamefont
  {de~Wit}}, \bibinfo {author} {\bibfnamefont {M.}~\bibnamefont {Fruchart}},
  \bibinfo {author} {\bibfnamefont {T.}~\bibnamefont {Khain}}, \bibinfo
  {author} {\bibfnamefont {F.}~\bibnamefont {Toschi}},\ and\ \bibinfo {author}
  {\bibfnamefont {V.}~\bibnamefont {Vitelli}},\ }\bibfield  {title} {\enquote
  {\bibinfo {title} {Pattern formation by turbulent cascades},}\ }\href
  {https://doi.org/https://doi.org/10.1038/s41586-024-07074-z} {\bibfield
  {journal} {\bibinfo  {journal} {Nature}\ }\textbf {\bibinfo {volume} {627}},\
  \bibinfo {pages} {515--521} (\bibinfo {year} {2024})}\BibitemShut {NoStop}%
\bibitem [{\citenamefont {Chen}\ \emph
  {et~al.}(2024{\natexlab{b}})\citenamefont {Chen}, \citenamefont {de~Wit},
  \citenamefont {Fruchart}, \citenamefont {Toschi},\ and\ \citenamefont
  {Vitelli}}]{chen2024odd}%
  \BibitemOpen
  \bibfield  {author} {\bibinfo {author} {\bibfnamefont {S.}~\bibnamefont
  {Chen}}, \bibinfo {author} {\bibfnamefont {X.~M.}\ \bibnamefont {de~Wit}},
  \bibinfo {author} {\bibfnamefont {M.}~\bibnamefont {Fruchart}}, \bibinfo
  {author} {\bibfnamefont {F.}~\bibnamefont {Toschi}},\ and\ \bibinfo {author}
  {\bibfnamefont {V.}~\bibnamefont {Vitelli}},\ }\bibfield  {title} {\enquote
  {\bibinfo {title} {Odd viscosity suppresses intermittency in direct turbulent
  cascades},}\ }\href
  {https://doi.org/https://doi.org/10.1103/PhysRevLett.133.144002} {\bibfield
  {journal} {\bibinfo  {journal} {Phys. Rev. Lett.}\ }\textbf {\bibinfo
  {volume} {133}},\ \bibinfo {pages} {144002} (\bibinfo {year}
  {2024}{\natexlab{b}})}\BibitemShut {NoStop}%
\bibitem [{\citenamefont {Hosaka}, \citenamefont {Komura},\ and\ \citenamefont
  {Andelman}(2021{\natexlab{a}})}]{hosaka2021nonreciprocal}%
  \BibitemOpen
  \bibfield  {author} {\bibinfo {author} {\bibfnamefont {Y.}~\bibnamefont
  {Hosaka}}, \bibinfo {author} {\bibfnamefont {S.}~\bibnamefont {Komura}},\
  and\ \bibinfo {author} {\bibfnamefont {D.}~\bibnamefont {Andelman}},\
  }\bibfield  {title} {\enquote {\bibinfo {title} {Nonreciprocal response of a
  two-dimensional fluid with odd viscosity},}\ }\href
  {https://doi.org/10.1103/PhysRevE.103.042610} {\bibfield  {journal} {\bibinfo
   {journal} {Phys. Rev. E}\ }\textbf {\bibinfo {volume} {103}},\ \bibinfo
  {pages} {042610} (\bibinfo {year} {2021}{\natexlab{a}})}\BibitemShut
  {NoStop}%
\bibitem [{\citenamefont {Hosaka}, \citenamefont {Komura},\ and\ \citenamefont
  {Andelman}(2021{\natexlab{b}})}]{hosaka2021hydrodynamic}%
  \BibitemOpen
  \bibfield  {author} {\bibinfo {author} {\bibfnamefont {Y.}~\bibnamefont
  {Hosaka}}, \bibinfo {author} {\bibfnamefont {S.}~\bibnamefont {Komura}},\
  and\ \bibinfo {author} {\bibfnamefont {D.}~\bibnamefont {Andelman}},\
  }\bibfield  {title} {\enquote {\bibinfo {title} {Hydrodynamic lift of a
  two-dimensional liquid domain with odd viscosity},}\ }\href
  {https://doi.org/10.1103/PhysRevE.104.064613} {\bibfield  {journal} {\bibinfo
   {journal} {Phys. Rev. E}\ }\textbf {\bibinfo {volume} {104}},\ \bibinfo
  {pages} {064613} (\bibinfo {year} {2021}{\natexlab{b}})}\BibitemShut
  {NoStop}%
\bibitem [{\citenamefont {Lier}\ \emph {et~al.}(2023)\citenamefont {Lier},
  \citenamefont {Duclut}, \citenamefont {Bo}, \citenamefont {Armas},
  \citenamefont {J{\"u}licher},\ and\ \citenamefont
  {Sur{\'o}wka}}]{lier2023lift}%
  \BibitemOpen
  \bibfield  {author} {\bibinfo {author} {\bibfnamefont {R.}~\bibnamefont
  {Lier}}, \bibinfo {author} {\bibfnamefont {C.}~\bibnamefont {Duclut}},
  \bibinfo {author} {\bibfnamefont {S.}~\bibnamefont {Bo}}, \bibinfo {author}
  {\bibfnamefont {J.}~\bibnamefont {Armas}}, \bibinfo {author} {\bibfnamefont
  {F.}~\bibnamefont {J{\"u}licher}},\ and\ \bibinfo {author} {\bibfnamefont
  {P.}~\bibnamefont {Sur{\'o}wka}},\ }\bibfield  {title} {\enquote {\bibinfo
  {title} {Lift force in odd compressible fluids},}\ }\href
  {https://doi.org/10.1103/PhysRevE.108.L023101} {\bibfield  {journal}
  {\bibinfo  {journal} {Phys. Rev. E}\ }\textbf {\bibinfo {volume} {108}},\
  \bibinfo {pages} {L023101} (\bibinfo {year} {2023})}\BibitemShut {NoStop}%
\bibitem [{\citenamefont {Everts}\ and\ \citenamefont
  {Cichocki}(2024)}]{everts2024dissipative}%
  \BibitemOpen
  \bibfield  {author} {\bibinfo {author} {\bibfnamefont {J.~C.}\ \bibnamefont
  {Everts}}\ and\ \bibinfo {author} {\bibfnamefont {B.}~\bibnamefont
  {Cichocki}},\ }\bibfield  {title} {\enquote {\bibinfo {title} {Dissipative
  effects in odd viscous {S}tokes flow around a single sphere},}\ }\href
  {https://doi.org/https://doi.org/10.1103/PhysRevLett.132.218303} {\bibfield
  {journal} {\bibinfo  {journal} {Phys. Rev. Lett.}\ }\textbf {\bibinfo
  {volume} {132}},\ \bibinfo {pages} {218303} (\bibinfo {year}
  {2024})}\BibitemShut {NoStop}%
\bibitem [{\citenamefont {Lier}(2024)}]{lier2024slip}%
  \BibitemOpen
  \bibfield  {author} {\bibinfo {author} {\bibfnamefont {R.}~\bibnamefont
  {Lier}},\ }\bibfield  {title} {\enquote {\bibinfo {title} {Slip-induced odd
  viscous flow past a cylinder},}\ }\href
  {https://doi.org/https://doi.org/10.1103/PhysRevFluids.9.094101} {\bibfield
  {journal} {\bibinfo  {journal} {Phys. Rev. Fluids}\ }\textbf {\bibinfo
  {volume} {9}},\ \bibinfo {pages} {094101} (\bibinfo {year}
  {2024})}\BibitemShut {NoStop}%
\bibitem [{\citenamefont {Khain}\ \emph {et~al.}(2024)\citenamefont {Khain},
  \citenamefont {Fruchart}, \citenamefont {Scheibner}, \citenamefont {Witten},\
  and\ \citenamefont {Vitelli}}]{khain2024trading}%
  \BibitemOpen
  \bibfield  {author} {\bibinfo {author} {\bibfnamefont {T.}~\bibnamefont
  {Khain}}, \bibinfo {author} {\bibfnamefont {M.}~\bibnamefont {Fruchart}},
  \bibinfo {author} {\bibfnamefont {C.}~\bibnamefont {Scheibner}}, \bibinfo
  {author} {\bibfnamefont {T.~A.}\ \bibnamefont {Witten}},\ and\ \bibinfo
  {author} {\bibfnamefont {V.}~\bibnamefont {Vitelli}},\ }\bibfield  {title}
  {\enquote {\bibinfo {title} {Trading particle shape with fluid symmetry: on
  the mobility matrix in 3-{D} chiral fluids},}\ }\href
  {https://doi.org/https://doi.org/10.1017/jfm.2024.535} {\bibfield  {journal}
  {\bibinfo  {journal} {J. Fluid Mech.}\ }\textbf {\bibinfo {volume} {992}},\
  \bibinfo {pages} {A5} (\bibinfo {year} {2024})}\BibitemShut {NoStop}%
\bibitem [{\citenamefont {Reichhardt}\ and\ \citenamefont
  {Reichhardt}(2022)}]{reichhardt2022}%
  \BibitemOpen
  \bibfield  {author} {\bibinfo {author} {\bibfnamefont {C.~J.~O.}\
  \bibnamefont {Reichhardt}}\ and\ \bibinfo {author} {\bibfnamefont
  {C.}~\bibnamefont {Reichhardt}},\ }\bibfield  {title} {\enquote {\bibinfo
  {title} {Active rheology in odd-viscosity systems},}\ }\href
  {https://doi.org/10.1209/0295-5075/ac2adc} {\bibfield  {journal} {\bibinfo
  {journal} {EPL}\ }\textbf {\bibinfo {volume} {137}},\ \bibinfo {pages}
  {66004} (\bibinfo {year} {2022})}\BibitemShut {NoStop}%
\bibitem [{\citenamefont {Lou}\ \emph {et~al.}(2022)\citenamefont {Lou},
  \citenamefont {Yang}, \citenamefont {Ding}, \citenamefont {Liu},
  \citenamefont {Chen}, \citenamefont {Zhou}, \citenamefont {Ye}, \citenamefont
  {Podgornik},\ and\ \citenamefont {Yang}}]{lou2022odd}%
  \BibitemOpen
  \bibfield  {author} {\bibinfo {author} {\bibfnamefont {X.}~\bibnamefont
  {Lou}}, \bibinfo {author} {\bibfnamefont {Q.}~\bibnamefont {Yang}}, \bibinfo
  {author} {\bibfnamefont {Y.}~\bibnamefont {Ding}}, \bibinfo {author}
  {\bibfnamefont {P.}~\bibnamefont {Liu}}, \bibinfo {author} {\bibfnamefont
  {K.}~\bibnamefont {Chen}}, \bibinfo {author} {\bibfnamefont {X.}~\bibnamefont
  {Zhou}}, \bibinfo {author} {\bibfnamefont {F.}~\bibnamefont {Ye}}, \bibinfo
  {author} {\bibfnamefont {R.}~\bibnamefont {Podgornik}},\ and\ \bibinfo
  {author} {\bibfnamefont {M.}~\bibnamefont {Yang}},\ }\bibfield  {title}
  {\enquote {\bibinfo {title} {Odd viscosity-induced {H}all-like transport of
  an active chiral fluid},}\ }\href {https://doi.org/10.1073/pnas.2201279119}
  {\bibfield  {journal} {\bibinfo  {journal} {Proc. Natl. Acad. Sci. U.S.A.}\
  }\textbf {\bibinfo {volume} {119}},\ \bibinfo {pages} {e2201279119} (\bibinfo
  {year} {2022})}\BibitemShut {NoStop}%
\bibitem [{\citenamefont {Lapa}\ and\ \citenamefont {Hughes}(2014)}]{lapa2014}%
  \BibitemOpen
  \bibfield  {author} {\bibinfo {author} {\bibfnamefont {M.~F.}\ \bibnamefont
  {Lapa}}\ and\ \bibinfo {author} {\bibfnamefont {T.~L.}\ \bibnamefont
  {Hughes}},\ }\bibfield  {title} {\enquote {\bibinfo {title} {Swimming at low
  {Reynolds} number in fluids with odd, or {Hall}, viscosity},}\ }\href
  {https://doi.org/10.1103/PhysRevE.89.043019} {\bibfield  {journal} {\bibinfo
  {journal} {Phys. Rev. E}\ }\textbf {\bibinfo {volume} {89}},\ \bibinfo
  {pages} {043019} (\bibinfo {year} {2014})}\BibitemShut {NoStop}%
\bibitem [{\citenamefont {Hosaka}, \citenamefont {Golestanian},\ and\
  \citenamefont {Vilfan}(2023)}]{hosaka2023lorentz}%
  \BibitemOpen
  \bibfield  {author} {\bibinfo {author} {\bibfnamefont {Y.}~\bibnamefont
  {Hosaka}}, \bibinfo {author} {\bibfnamefont {R.}~\bibnamefont
  {Golestanian}},\ and\ \bibinfo {author} {\bibfnamefont {A.}~\bibnamefont
  {Vilfan}},\ }\bibfield  {title} {\enquote {\bibinfo {title} {Lorentz
  reciprocal theorem in fluids with odd viscosity},}\ }\href
  {https://doi.org/10.1103/PhysRevLett.131.178303} {\bibfield  {journal}
  {\bibinfo  {journal} {Phys. Rev. Lett.}\ }\textbf {\bibinfo {volume} {131}},\
  \bibinfo {pages} {178303} (\bibinfo {year} {2023})}\BibitemShut {NoStop}%
\bibitem [{\citenamefont {Hosaka}\ \emph {et~al.}(2024)\citenamefont {Hosaka},
  \citenamefont {Chatzittofi}, \citenamefont {Golestanian},\ and\ \citenamefont
  {Vilfan}}]{hosaka2024chirotactic}%
  \BibitemOpen
  \bibfield  {author} {\bibinfo {author} {\bibfnamefont {Y.}~\bibnamefont
  {Hosaka}}, \bibinfo {author} {\bibfnamefont {M.}~\bibnamefont {Chatzittofi}},
  \bibinfo {author} {\bibfnamefont {R.}~\bibnamefont {Golestanian}},\ and\
  \bibinfo {author} {\bibfnamefont {A.}~\bibnamefont {Vilfan}},\ }\bibfield
  {title} {\enquote {\bibinfo {title} {Chirotactic response of microswimmers in
  fluids with odd viscosity},}\ }\href
  {https://doi.org/https://doi.org/10.1103/PhysRevResearch.6.L032044}
  {\bibfield  {journal} {\bibinfo  {journal} {Phys. Rev. Research}\ }\textbf
  {\bibinfo {volume} {6}},\ \bibinfo {pages} {L032044} (\bibinfo {year}
  {2024})}\BibitemShut {NoStop}%
\bibitem [{\citenamefont {Khain}\ \emph {et~al.}(2022)\citenamefont {Khain},
  \citenamefont {Scheibner}, \citenamefont {Fruchart},\ and\ \citenamefont
  {Vitelli}}]{khain2022}%
  \BibitemOpen
  \bibfield  {author} {\bibinfo {author} {\bibfnamefont {T.}~\bibnamefont
  {Khain}}, \bibinfo {author} {\bibfnamefont {C.}~\bibnamefont {Scheibner}},
  \bibinfo {author} {\bibfnamefont {M.}~\bibnamefont {Fruchart}},\ and\
  \bibinfo {author} {\bibfnamefont {V.}~\bibnamefont {Vitelli}},\ }\bibfield
  {title} {\enquote {\bibinfo {title} {Stokes flows in three-dimensional fluids
  with odd and parity-violating viscosities},}\ }\href
  {https://doi.org/10.1017/jfm.2021.1079} {\bibfield  {journal} {\bibinfo
  {journal} {J. Fluid Mech.}\ }\textbf {\bibinfo {volume} {934}},\ \bibinfo
  {pages} {A23} (\bibinfo {year} {2022})}\BibitemShut {NoStop}%
\bibitem [{\citenamefont {Ganeshan}\ and\ \citenamefont
  {Abanov}(2017)}]{ganeshan2017}%
  \BibitemOpen
  \bibfield  {author} {\bibinfo {author} {\bibfnamefont {S.}~\bibnamefont
  {Ganeshan}}\ and\ \bibinfo {author} {\bibfnamefont {A.~G.}\ \bibnamefont
  {Abanov}},\ }\bibfield  {title} {\enquote {\bibinfo {title} {Odd viscosity in
  two-dimensional incompressible fluids},}\ }\href
  {https://doi.org/10.1103/PhysRevFluids.2.094101} {\bibfield  {journal}
  {\bibinfo  {journal} {Phys. Rev. Fluids}\ }\textbf {\bibinfo {volume} {2}},\
  \bibinfo {pages} {094101} (\bibinfo {year} {2017})}\BibitemShut {NoStop}%
\bibitem [{\citenamefont {Jia}, \citenamefont {Irvine},\ and\ \citenamefont
  {Shelley}(2022)}]{jia2022incompressible}%
  \BibitemOpen
  \bibfield  {author} {\bibinfo {author} {\bibfnamefont {L.~L.}\ \bibnamefont
  {Jia}}, \bibinfo {author} {\bibfnamefont {W.~T.~M.}\ \bibnamefont {Irvine}},\
  and\ \bibinfo {author} {\bibfnamefont {M.~J.}\ \bibnamefont {Shelley}},\
  }\bibfield  {title} {\enquote {\bibinfo {title} {Incompressible active phases
  at an interface. {P}art 1. {F}ormulation and axisymmetric odd flows},}\
  }\href {https://doi.org/https://doi.org/10.1017/jfm.2022.856} {\bibfield
  {journal} {\bibinfo  {journal} {J. Fluid Mech.}\ }\textbf {\bibinfo {volume}
  {951}},\ \bibinfo {pages} {A36} (\bibinfo {year} {2022})}\BibitemShut
  {NoStop}%
\bibitem [{\citenamefont {Epstein}\ and\ \citenamefont
  {Mandadapu}(2020)}]{epstein2020}%
  \BibitemOpen
  \bibfield  {author} {\bibinfo {author} {\bibfnamefont {J.~M.}\ \bibnamefont
  {Epstein}}\ and\ \bibinfo {author} {\bibfnamefont {K.~K.}\ \bibnamefont
  {Mandadapu}},\ }\bibfield  {title} {\enquote {\bibinfo {title} {Time-reversal
  symmetry breaking in two-dimensional nonequilibrium viscous fluids},}\ }\href
  {https://doi.org/https://doi.org/10.1103/PhysRevE.101.052614} {\bibfield
  {journal} {\bibinfo  {journal} {Phys. Rev. E}\ }\textbf {\bibinfo {volume}
  {101}},\ \bibinfo {pages} {052614} (\bibinfo {year} {2020})}\BibitemShut
  {NoStop}%
\bibitem [{\citenamefont {Souslov}, \citenamefont {Gromov},\ and\ \citenamefont
  {Vitelli}(2020)}]{souslov2020}%
  \BibitemOpen
  \bibfield  {author} {\bibinfo {author} {\bibfnamefont {A.}~\bibnamefont
  {Souslov}}, \bibinfo {author} {\bibfnamefont {A.}~\bibnamefont {Gromov}},\
  and\ \bibinfo {author} {\bibfnamefont {V.}~\bibnamefont {Vitelli}},\
  }\bibfield  {title} {\enquote {\bibinfo {title} {Anisotropic odd viscosity
  via a time-modulated drive},}\ }\href
  {https://doi.org/https://doi.org/10.1103/PhysRevE.101.052606} {\bibfield
  {journal} {\bibinfo  {journal} {Phys. Rev. E}\ }\textbf {\bibinfo {volume}
  {101}},\ \bibinfo {pages} {052606} (\bibinfo {year} {2020})}\BibitemShut
  {NoStop}%
\bibitem [{\citenamefont {Barentin}\ \emph {et~al.}(1999)\citenamefont
  {Barentin}, \citenamefont {Ybert}, \citenamefont {di~Meglio},\ and\
  \citenamefont {Joanny}}]{barentin1999}%
  \BibitemOpen
  \bibfield  {author} {\bibinfo {author} {\bibfnamefont {C.}~\bibnamefont
  {Barentin}}, \bibinfo {author} {\bibfnamefont {C.}~\bibnamefont {Ybert}},
  \bibinfo {author} {\bibfnamefont {J.-M.}\ \bibnamefont {di~Meglio}},\ and\
  \bibinfo {author} {\bibfnamefont {J.-F.}\ \bibnamefont {Joanny}},\ }\bibfield
   {title} {\enquote {\bibinfo {title} {Surface shear viscosity of {G}ibbs and
  {L}angmuir monolayers},}\ }\href {https://doi.org/10.1017/S0022112099006321}
  {\bibfield  {journal} {\bibinfo  {journal} {J. Fluid Mech.}\ }\textbf
  {\bibinfo {volume} {397}},\ \bibinfo {pages} {331--349} (\bibinfo {year}
  {1999})}\BibitemShut {NoStop}%
\bibitem [{\citenamefont {Duclut}\ \emph {et~al.}(2024)\citenamefont {Duclut},
  \citenamefont {Bo}, \citenamefont {Lier}, \citenamefont {Armas},
  \citenamefont {Sur{\'o}wka},\ and\ \citenamefont
  {J{\"u}licher}}]{duclut2024probe}%
  \BibitemOpen
  \bibfield  {author} {\bibinfo {author} {\bibfnamefont {C.}~\bibnamefont
  {Duclut}}, \bibinfo {author} {\bibfnamefont {S.}~\bibnamefont {Bo}}, \bibinfo
  {author} {\bibfnamefont {R.}~\bibnamefont {Lier}}, \bibinfo {author}
  {\bibfnamefont {J.}~\bibnamefont {Armas}}, \bibinfo {author} {\bibfnamefont
  {P.}~\bibnamefont {Sur{\'o}wka}},\ and\ \bibinfo {author} {\bibfnamefont
  {F.}~\bibnamefont {J{\"u}licher}},\ }\bibfield  {title} {\enquote {\bibinfo
  {title} {Probe particles in odd active viscoelastic fluids: How activity and
  dissipation determine linear stability},}\ }\href
  {https://doi.org/https://doi.org/10.1103/PhysRevE.109.044126} {\bibfield
  {journal} {\bibinfo  {journal} {Phys. Rev. E}\ }\textbf {\bibinfo {volume}
  {109}},\ \bibinfo {pages} {044126} (\bibinfo {year} {2024})}\BibitemShut
  {NoStop}%
\bibitem [{\citenamefont {Hosaka}, \citenamefont {Andelman},\ and\
  \citenamefont {Komura}(2023)}]{hosaka2023pair}%
  \BibitemOpen
  \bibfield  {author} {\bibinfo {author} {\bibfnamefont {Y.}~\bibnamefont
  {Hosaka}}, \bibinfo {author} {\bibfnamefont {D.}~\bibnamefont {Andelman}},\
  and\ \bibinfo {author} {\bibfnamefont {S.}~\bibnamefont {Komura}},\
  }\bibfield  {title} {\enquote {\bibinfo {title} {Pair dynamics of active
  force dipoles in an odd-viscous fluid},}\ }\href
  {https://doi.org/10.1140/epje/s10189-023-00265-y} {\bibfield  {journal}
  {\bibinfo  {journal} {Eur. Phys. J. E}\ }\textbf {\bibinfo {volume} {46}},\
  \bibinfo {pages} {18} (\bibinfo {year} {2023})}\BibitemShut {NoStop}%
\bibitem [{\citenamefont {Hosaka}, \citenamefont {Golestanian},\ and\
  \citenamefont {Daddi-Moussa-Ider}(2023)}]{hosaka2023hydrodynamics}%
  \BibitemOpen
  \bibfield  {author} {\bibinfo {author} {\bibfnamefont {Y.}~\bibnamefont
  {Hosaka}}, \bibinfo {author} {\bibfnamefont {R.}~\bibnamefont
  {Golestanian}},\ and\ \bibinfo {author} {\bibfnamefont {A.}~\bibnamefont
  {Daddi-Moussa-Ider}},\ }\bibfield  {title} {\enquote {\bibinfo {title}
  {Hydrodynamics of an odd active surfer in a chiral fluid},}\ }\href
  {https://doi.org/10.1088/1367-2630/aceea4} {\bibfield  {journal} {\bibinfo
  {journal} {New J. Phys.}\ }\textbf {\bibinfo {volume} {25}},\ \bibinfo
  {pages} {083046} (\bibinfo {year} {2023})}\BibitemShut {NoStop}%
\bibitem [{\citenamefont {Elfring}, \citenamefont {Leal},\ and\ \citenamefont
  {Squires}(2016)}]{elfring2016surface}%
  \BibitemOpen
  \bibfield  {author} {\bibinfo {author} {\bibfnamefont {G.~J.}\ \bibnamefont
  {Elfring}}, \bibinfo {author} {\bibfnamefont {L.~G.}\ \bibnamefont {Leal}},\
  and\ \bibinfo {author} {\bibfnamefont {T.~M.}\ \bibnamefont {Squires}},\
  }\bibfield  {title} {\enquote {\bibinfo {title} {Surface viscosity and
  {Marangoni} stresses at surfactant laden interfaces},}\ }\href
  {https://doi.org/https://doi.org/10.1017/jfm.2016.96} {\bibfield  {journal}
  {\bibinfo  {journal} {J. Fluid Mech.}\ }\textbf {\bibinfo {volume} {792}},\
  \bibinfo {pages} {712--739} (\bibinfo {year} {2016})}\BibitemShut {NoStop}%
\bibitem [{\citenamefont {Felderhof}(2006)}]{felderhof2006dynamics}%
  \BibitemOpen
  \bibfield  {author} {\bibinfo {author} {\bibfnamefont {B.}~\bibnamefont
  {Felderhof}},\ }\bibfield  {title} {\enquote {\bibinfo {title} {Dynamics of
  an interface with adsorption layer between two fluids},}\ }\href
  {https://doi.org/10.1063/1.2372460} {\bibfield  {journal} {\bibinfo
  {journal} {Phys. Fluids}\ }\textbf {\bibinfo {volume} {18}} (\bibinfo {year}
  {2006})}\BibitemShut {NoStop}%
\bibitem [{\citenamefont {Daddi-Moussa-Ider}\ and\ \citenamefont
  {Gekle}(2016)}]{daddi2016hydrodynamic}%
  \BibitemOpen
  \bibfield  {author} {\bibinfo {author} {\bibfnamefont {A.}~\bibnamefont
  {Daddi-Moussa-Ider}}\ and\ \bibinfo {author} {\bibfnamefont {S.}~\bibnamefont
  {Gekle}},\ }\bibfield  {title} {\enquote {\bibinfo {title} {Hydrodynamic
  interaction between particles near elastic interfaces},}\ }\href
  {https://doi.org/10.1063/1.4955099} {\bibfield  {journal} {\bibinfo
  {journal} {J. Chem. Phys.}\ }\textbf {\bibinfo {volume} {145}} (\bibinfo
  {year} {2016})}\BibitemShut {NoStop}%
\bibitem [{\citenamefont {Daddi-Moussa-Ider}\ and\ \citenamefont
  {Gekle}(2018)}]{daddi2018brownian}%
  \BibitemOpen
  \bibfield  {author} {\bibinfo {author} {\bibfnamefont {A.}~\bibnamefont
  {Daddi-Moussa-Ider}}\ and\ \bibinfo {author} {\bibfnamefont {S.}~\bibnamefont
  {Gekle}},\ }\bibfield  {title} {\enquote {\bibinfo {title} {Brownian motion
  near an elastic cell membrane: A theoretical study},}\ }\href
  {https://doi.org/https://doi.org/10.1140/epje/i2018-11627-6} {\bibfield
  {journal} {\bibinfo  {journal} {Eur. Phys. J. E}\ }\textbf {\bibinfo {volume}
  {41}},\ \bibinfo {pages} {1--13} (\bibinfo {year} {2018})}\BibitemShut
  {NoStop}%
\bibitem [{\citenamefont {Daddi-Moussa-Ider}\ \emph {et~al.}(2018)\citenamefont
  {Daddi-Moussa-Ider}, \citenamefont {Lisicki}, \citenamefont {Gekle},
  \citenamefont {Menzel},\ and\ \citenamefont
  {L{\"o}wen}}]{daddi2018hydrodynamic}%
  \BibitemOpen
  \bibfield  {author} {\bibinfo {author} {\bibfnamefont {A.}~\bibnamefont
  {Daddi-Moussa-Ider}}, \bibinfo {author} {\bibfnamefont {M.}~\bibnamefont
  {Lisicki}}, \bibinfo {author} {\bibfnamefont {S.}~\bibnamefont {Gekle}},
  \bibinfo {author} {\bibfnamefont {A.~M.}\ \bibnamefont {Menzel}},\ and\
  \bibinfo {author} {\bibfnamefont {H.}~\bibnamefont {L{\"o}wen}},\ }\bibfield
  {title} {\enquote {\bibinfo {title} {Hydrodynamic coupling and rotational
  mobilities near planar elastic membranes},}\ }\href
  {https://doi.org/10.1063/1.5032304} {\bibfield  {journal} {\bibinfo
  {journal} {J. Chem. Phys.}\ }\textbf {\bibinfo {volume} {149}} (\bibinfo
  {year} {2018})}\BibitemShut {NoStop}%
\bibitem [{\citenamefont {Daddi-Moussa-Ider}\ \emph
  {et~al.}(2024{\natexlab{a}})\citenamefont {Daddi-Moussa-Ider}, \citenamefont
  {Tjhung}, \citenamefont {Richter},\ and\ \citenamefont
  {Menzel}}]{Daddi-Moussa-Ider_2024_JPCM}%
  \BibitemOpen
  \bibfield  {author} {\bibinfo {author} {\bibfnamefont {A.}~\bibnamefont
  {Daddi-Moussa-Ider}}, \bibinfo {author} {\bibfnamefont {E.}~\bibnamefont
  {Tjhung}}, \bibinfo {author} {\bibfnamefont {T.}~\bibnamefont {Richter}},\
  and\ \bibinfo {author} {\bibfnamefont {A.~M.}\ \bibnamefont {Menzel}},\
  }\bibfield  {title} {\enquote {\bibinfo {title} {Hydrodynamics of a disk in a
  thin film of weakly nematic fluid subject to linear friction},}\ }\href
  {https://doi.org/10.1088/1361-648X/ad65ad} {\bibfield  {journal} {\bibinfo
  {journal} {J. Phys.: Condens. Matter}\ }\textbf {\bibinfo {volume} {36}},\
  \bibinfo {pages} {445101} (\bibinfo {year} {2024}{\natexlab{a}})}\BibitemShut
  {NoStop}%
\bibitem [{\citenamefont {Daddi-Moussa-Ider}\ \emph
  {et~al.}(2024{\natexlab{b}})\citenamefont {Daddi-Moussa-Ider}, \citenamefont
  {Tjhung}, \citenamefont {Pradas}, \citenamefont {Richter},\ and\
  \citenamefont {Menzel}}]{daddi2024rotational}%
  \BibitemOpen
  \bibfield  {author} {\bibinfo {author} {\bibfnamefont {A.}~\bibnamefont
  {Daddi-Moussa-Ider}}, \bibinfo {author} {\bibfnamefont {E.}~\bibnamefont
  {Tjhung}}, \bibinfo {author} {\bibfnamefont {M.}~\bibnamefont {Pradas}},
  \bibinfo {author} {\bibfnamefont {T.}~\bibnamefont {Richter}},\ and\ \bibinfo
  {author} {\bibfnamefont {A.~M.}\ \bibnamefont {Menzel}},\ }\bibfield  {title}
  {\enquote {\bibinfo {title} {Rotational dynamics of a disk in a thin film of
  weakly nematic fluid subject to linear friction},}\ }\href
  {https://doi.org/https://doi.org/10.1140/epje/s10189-024-00452-5} {\bibfield
  {journal} {\bibinfo  {journal} {Eur. Phys. J. E}\ }\textbf {\bibinfo {volume}
  {47}},\ \bibinfo {pages} {58} (\bibinfo {year}
  {2024}{\natexlab{b}})}\BibitemShut {NoStop}%
\bibitem [{\citenamefont {Daddi-Moussa-Ider}, \citenamefont {Golestanian},\
  and\ \citenamefont {Vilfan}(2024)}]{daddi2024hydrodynamic_JFM}%
  \BibitemOpen
  \bibfield  {author} {\bibinfo {author} {\bibfnamefont {A.}~\bibnamefont
  {Daddi-Moussa-Ider}}, \bibinfo {author} {\bibfnamefont {R.}~\bibnamefont
  {Golestanian}},\ and\ \bibinfo {author} {\bibfnamefont {A.}~\bibnamefont
  {Vilfan}},\ }\bibfield  {title} {\enquote {\bibinfo {title} {Hydrodynamic
  efficiency limit on a {Marangoni} surfer},}\ }\href
  {https://doi.org/https://doi.org/10.1017/jfm.2024.363} {\bibfield  {journal}
  {\bibinfo  {journal} {J. Fluid Mech.}\ }\textbf {\bibinfo {volume} {986}},\
  \bibinfo {pages} {A32} (\bibinfo {year} {2024})}\BibitemShut {NoStop}%
\bibitem [{\citenamefont {Baddour}(2011)}]{baddour2011two}%
  \BibitemOpen
  \bibfield  {author} {\bibinfo {author} {\bibfnamefont {N.}~\bibnamefont
  {Baddour}},\ }\bibfield  {title} {\enquote {\bibinfo {title} {Two-dimensional
  {Fourier} transforms in polar coordinates},}\ }in\ \href
  {https://doi.org/https://doi.org/10.1016/B978-0-12-385861-0.00001-4} {\emph
  {\bibinfo {booktitle} {Adv. Imaging Electron Phys.}}},\ Vol.\ \bibinfo
  {volume} {165}\ (\bibinfo  {publisher} {Elsevier},\ \bibinfo {year} {2011})\
  pp.\ \bibinfo {pages} {1--45}\BibitemShut {NoStop}%
\bibitem [{\citenamefont {Abramowitz}\ and\ \citenamefont
  {Stegun}(2000)}]{abramowitz2000handbook}%
  \BibitemOpen
  \bibfield  {author} {\bibinfo {author} {\bibfnamefont {M.}~\bibnamefont
  {Abramowitz}}\ and\ \bibinfo {author} {\bibfnamefont {I.~A.}\ \bibnamefont
  {Stegun}},\ }\href@noop {} {\emph {\bibinfo {title} {{Handbook of
  Mathematical Functions with Formulas, Graphs, and Mathematical Tables}}}},\
  Vol.~\bibinfo {volume} {55}\ (\bibinfo  {publisher} {Dover Publications Inc,
  New York},\ \bibinfo {year} {2000})\BibitemShut {NoStop}%
\bibitem [{\citenamefont {Saffman}\ and\ \citenamefont
  {Delbr{\"u}ck}(1975)}]{saffman1975}%
  \BibitemOpen
  \bibfield  {author} {\bibinfo {author} {\bibfnamefont {P.~G.}\ \bibnamefont
  {Saffman}}\ and\ \bibinfo {author} {\bibfnamefont {M.}~\bibnamefont
  {Delbr{\"u}ck}},\ }\bibfield  {title} {\enquote {\bibinfo {title} {Brownian
  motion in biological membranes},}\ }\href
  {https://doi.org/https://doi.org/10.1073/pnas.72.8.3111} {\bibfield
  {journal} {\bibinfo  {journal} {Proc. Natl. Acad. Sci. U.S.A.}\ }\textbf
  {\bibinfo {volume} {72}},\ \bibinfo {pages} {3111--3113} (\bibinfo {year}
  {1975})}\BibitemShut {NoStop}%
\bibitem [{\citenamefont {Saffman}(1976)}]{saffman1976}%
  \BibitemOpen
  \bibfield  {author} {\bibinfo {author} {\bibfnamefont {P.~G.}\ \bibnamefont
  {Saffman}},\ }\bibfield  {title} {\enquote {\bibinfo {title} {Brownian motion
  in thin sheets of viscous fluid},}\ }\href
  {https://doi.org/https://doi.org/10.1017/S0022112076001511} {\bibfield
  {journal} {\bibinfo  {journal} {J. Fluid Mech.}\ }\textbf {\bibinfo {volume}
  {73}},\ \bibinfo {pages} {593--602} (\bibinfo {year} {1976})}\BibitemShut
  {NoStop}%
\bibitem [{\citenamefont {Evans}\ and\ \citenamefont
  {Sackmann}(1988)}]{evans1988}%
  \BibitemOpen
  \bibfield  {author} {\bibinfo {author} {\bibfnamefont {E.}~\bibnamefont
  {Evans}}\ and\ \bibinfo {author} {\bibfnamefont {E.}~\bibnamefont
  {Sackmann}},\ }\bibfield  {title} {\enquote {\bibinfo {title} {Translational
  and rotational drag coefficients for a disk moving in a liquid membrane
  associated with a rigid substrate},}\ }\href
  {https://doi.org/https://doi.org/10.1017/S0022112088003106} {\bibfield
  {journal} {\bibinfo  {journal} {J. Fluid Mech.}\ }\textbf {\bibinfo {volume}
  {194}},\ \bibinfo {pages} {553--561} (\bibinfo {year} {1988})}\BibitemShut
  {NoStop}%
\bibitem [{\citenamefont {Diamant}(2009)}]{diamant2009hydrodynamic}%
  \BibitemOpen
  \bibfield  {author} {\bibinfo {author} {\bibfnamefont {H.}~\bibnamefont
  {Diamant}},\ }\bibfield  {title} {\enquote {\bibinfo {title} {Hydrodynamic
  interaction in confined geometries},}\ }\href
  {https://doi.org/https://doi.org/10.1143/JPSJ.78.041002} {\bibfield
  {journal} {\bibinfo  {journal} {Journal of the Physical Society of Japan}\
  }\textbf {\bibinfo {volume} {78}},\ \bibinfo {pages} {041002--041002}
  (\bibinfo {year} {2009})}\BibitemShut {NoStop}%
\end{thebibliography}
%

\end{document}